\documentclass[preprint,12pt]{elsarticle}

\usepackage{graphicx}
\usepackage{subcaption}
\usepackage[numbers]{natbib}
\usepackage{hyperref}
\usepackage{array}
\usepackage{booktabs}
\usepackage{multirow}
\usepackage{multicol}
\usepackage{float}
\usepackage{amsfonts}
\usepackage{amssymb}
\usepackage{amsmath}
\usepackage{algorithm}
\usepackage{algorithmic}
\usepackage{xcolor}
\usepackage{tikz}
\usetikzlibrary{arrows.meta,positioning}
\usepackage{amsthm}
\usepackage{xurl}

\newtheorem{theorem}{Theorem}
\newtheorem{lemma}{Lemma}

\newtheorem{definition}{Definition}

\long\def\ignore#1{}

\journal{European Journal of Operational Research}

\begin{document}

\begin{frontmatter}

\title{On  Analyzing the Conditions for Stability of Opportunistic Supply Chains Under Network Growth}

\author{Gurkirat Wadhwa}
\author{Priyank Sinha}

\affiliation{organization={IIT Bombay},
            addressline={Powai},
            city={Mumbai},
            postcode={400076},
            state={Maharashtra},
            country={India}}

\begin{abstract}
Even large firms such as Walmart, Apple, and Coca-Cola face persistent fluctuations in costs, demand, and raw material availability. These are not \textit{rare events} and cannot be evaluated using traditional disruption models focused on infrequent events. Instead, sustained volatility induces opportunistic behavior, as firms repeatedly reconfigure partners in absence of long-term contracts, often due to trust deficits. The resulting web of transient relationships forms opportunistic supply chains (OSCs).
To capture OSC evolution, we develop an integrated mathematical framework combining a Geometric Brownian Motion (GBM) model to represent stochastic price volatility, a Bayesian learning model to describe adaptive belief updates regarding partner reliability, and a Latent Order Logistic (LOLOG) network model for endogenous changes in network structure. This framework is implemented in an agent-based simulation to examine how volatility, trust, and network structure jointly shape SC resilience.
Our modeling approach identifies critical volatility threshold; a tipping point beyond which the network shifts from a stable, link-preserving regime to a fragmented regime marked by rapid relationship dissolution. We analytically establish monotonic effects of volatility on profitability, trust, and link activation; derive formal stability conditions and volatility-driven phase transitions, and show how these mechanisms shape node importance and procurement behavior. These theoretical mechanisms are illustrated through computational experiments reflecting industry behaviors in fast fashion, electronics, and perishables. Overall, our contribution is to develop an integrated GBM-Bayesian-LOLOG framework to analyze OSC stability and our model can be extended to other OSCs including humanitarian, pharmaceutical, and poultry networks.
\end{abstract}

\begin{graphicalabstract}
\end{graphicalabstract}

\begin{highlights}
\item dynamics of opportunistic supply chain modelled by Integrated GBM--Bayesian--LOLOG model
\item Network stability conditions under critical volatility threshold Stability theorem.
\item Simulation metrics quantify resilience under shocks across industries.
\end{highlights}


\end{frontmatter}

\section{Introduction}
Factories shut down overnight, suppliers disappear from digital marketplaces, and retailers rapidly reconfigure sourcing networks. Across sectors from fast fashion to semiconductors,modern supply chains (SCs) no longer resemble the stable, relational systems idealized in classical theory. Instead, they operate as volatile, adaptive networks that expand, fragment, and recombine in response to shocks, power asymmetries, and opportunistic behavior. In retail, firms such as Walmart and Costco leverage exclusivity agreements and slotting fees, contributing to high supplier churn \cite{foros2008slotting,richards2004}. In high-technology manufacturing, rigid contracting structures, exemplified by Apple’s dispute with GT Advanced Technologies, have precipitated supplier distress and bankruptcy under market turbulence \cite{gopinath2014,sec2019gt}.

These recurring disruptions reflect the rise of \emph{opportunistic supply chains} (OSCs): highly dynamic networks characterized by frequent partner switching, short-lived contracts, and incentives shaped more by volatility than cooperation. Classical SC theory emphasizes cooperative mechanisms such as \textit{buyback}, \textit{revenue-sharing}, and \textit{quantity-flexibility} contracts \cite{tsay2007}, yet such structures are difficult to sustain when uncertainty dominates. Trust erodes, contracting horizons shorten, transactional behavior intensifies, and \emph{wholesale-price contracts} \cite{cachon2003} prevail as networks continuously reconfigure.

Although OSCs are widely observed, most research examines uncertainty, opportunism, trust, or network formation in isolation \cite{loneragan2013,handley2012,chauhan2013}. Far fewer studies investigate how volatility, evolving trust, and endogenous restructuring jointly determine whether such systems persist or fragment. This motivates our central question: \emph{How do stochastic volatility, adaptive trust, and endogenous network restructuring together shape the evolution and stability of OSCs?}

We address this question through an integrated analytical framework combining GBM-based price dynamics, Bayesian trust updating, and LOLOG-based network evolution. We validate the resulting theoretical mechanisms using agent-based simulations calibrated to fast fashion, electronics, and perishable agricultural settings, explaining when opportunistic systems maintain structural persistence and when they disintegrate under sustained volatility. Section~\ref{sec_model} presents the framework, Section~\ref{sec:analysis} develops analytical results, Section~\ref{sec:simulations} reports simulation insights, and Section~\ref{sec:conclusion} concludes the work and lays future research direction.

\subsection*{Literature Review}
Research on SC dynamics provides extensive evidence on how uncertainty, opportunism, and adaptive behavior shape inter-firm relationships. Transaction cost and governance research studies show that when environments are volatile or poorly coordinated, firms often rely on short-term relationships. This increases the likelihood of coordination failures and exchange risks \cite{williamson1996,handley2012}. Trust and relational governance studies further demonstrate that turbulence, power asymmetries, and limited safeguards reduce relational continuity, particularly in outsourced and perishable contexts \cite{loneragan2013,lumineau2020}. At the same time, the resilience and agility literature suggests that strategic reconfiguration can strengthen responsiveness and mitigate performance loss when appropriately governed \cite{revilla2017}.

Network restructuring in SCs has been examined through models of stochastic network growth, preferential attachment, and opportunistic partner search, highlighting how structural evolution affects performance and vulnerability \cite{chauhan2013}. Statistical network approaches such as ERGM and LOLOG provide powerful tools to represent endogenous dependence, clustering, and relational turnover in evolving SCs \cite{fellows2019}. However, these network models generally capture link dynamics from a structural standpoint, while behavioral drivers such as evolving trust, changing profitability, and opportunistic incentives are often modeled separately, despite evidence that exchange hazards materially influence link survival \cite{handley2012}.

In parallel, volatility propagation and risk-focused resilience studies analyse how price shocks and uncertainty diffuse through supply networks and alter operational outcomes \cite{pilipovic1998,ivanov2014ripple}. Learning-based perspectives highlight that firms update beliefs and adapt behavior under uncertainty. Recent  works  have significantly advanced understanding of disruption propagation, network viability, and adaptive supply architectures under uncertainty \cite{ivanov2018ripple,ivanov2021viability,dolgui2020ripple,snyder2016}. Methodologically, mean-field and stochastic equilibrium approaches provide elegant system-level descriptions of interacting agents, typically under rational-expectation assumptions \cite{gast2012meanfield}.

Overall, existing works in literature do not jointly integrate: (i) stochastic
price volatility, (ii) Bayesian trust learning, and (iii) endogenous formation and dissolution of trading links, despite substantial empirical evidence that these mechanisms interact to determine opportunistic behavior and structural resilience . To address this gap, we develop an integrated framework that synthesizes GBM-based pricing, continuous Bayesian trust updating, and LOLOG-driven adaptive network evolution to characterize when OSCs remain stable, when they fragment, and how resilience emerges under volatility and bounded rationality.

\section{Model Formulation}\label{sec_model}
We develop an integrated framework that captures three behavioral mechanisms central to opportunistic supply chains (OSCs): stochastic pricing, adaptive trust, and endogenous network evolution. Prices follow \emph{Geometric Brownian Motion (GBM)}, reflecting continuous, non-negative, and volatile dynamics in short-horizon, weakly contracted markets. Inter-firm trust evolves through \emph{Bayesian learning}, as firms update beliefs about partner reliability from repeated, noisy outcomes under incomplete information. Network structure adjusts endogenously through a \emph{Latent-Order Logistic (LOLOG)} model, where profitability, trust, and structural incentives jointly govern link formation and dissolution.

\begin{figure}[h]
\centering

\begin{subfigure}[t]{0.45\columnwidth}
\centering
\includegraphics[width=\linewidth]{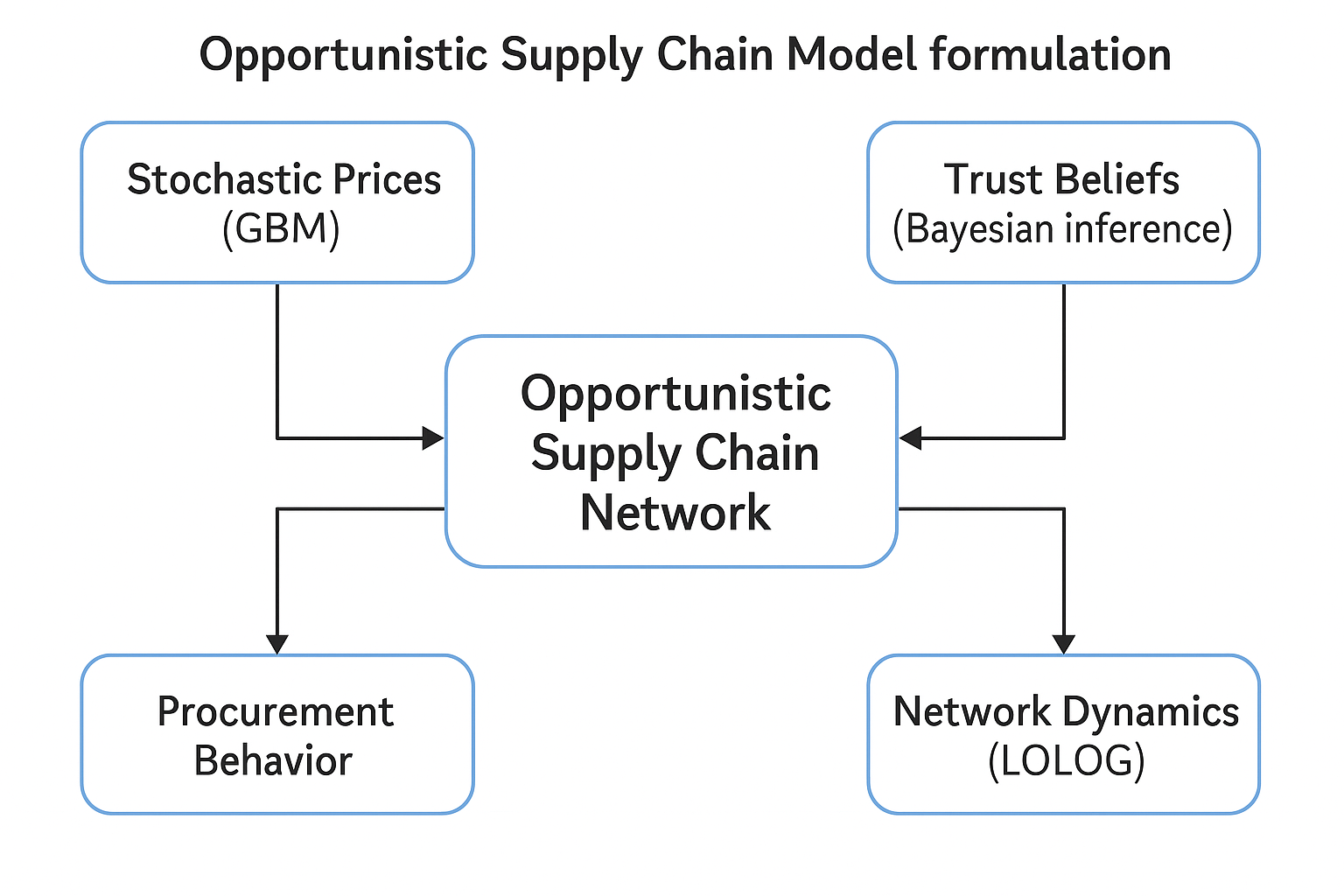}
\caption{Model formulation}
\label{fig:model}
\end{subfigure}
\hfill
\begin{subfigure}[t]{0.5\columnwidth}
\centering
\resizebox{\linewidth}{!}{%
\begin{tikzpicture}[
    node distance=1.4cm,
    rectangle,
    draw,
    rounded corners,
    align=center,
    minimum width=4.8cm,
    minimum height=1.0cm,
    >=Stealth
]

\node(gbm){Price Volatility \\ (GBM)};
\node(profit)[below of=gbm]{Profitability \& \\ Transaction Outcomes};
\node(trust)[below of=profit]{Trust Beliefs \\ (Bayesian Updating)};
\node(lolog)[below of=trust]{Link Formation / Dissolution \\ (LOLOG)};
\node(network)[below of=lolog]{Resulting Network Structure};

\draw[->, thick] (gbm) -- (profit);
\draw[->, thick] (profit) -- (trust);
\draw[->, thick] (trust) -- (lolog);
\draw[->, thick] (lolog) -- (network);

\draw[thick, ->]
   ([xshift=0.2cm]network.east)
   .. controls +(3cm,0cm) and +(3cm,0cm) ..
   node[midway, right]
   {\small Exposure to Risk\\ \& Incentives}
   ([xshift=0.2cm]gbm.east);

\end{tikzpicture}}
\caption{Behavioural process}
\label{fig:processflow}
\end{subfigure}

\caption{Integrated GBM--Bayesian--LOLOG framework and behavioural feedback process.}
\label{fig:combined}
\end{figure}

These mechanisms operate as a unified process: price shocks affect profitability, profitability shapes trust, and updated beliefs drive partner switching and network restructuring (Figure~\ref{fig:model}). The resulting network structure then influences how vulnerable firms are to future volatility. Highly concentrated or fragile networks amplify shocks, while more diversified networks help absorb them (Figure~\ref{fig:processflow}). In this multi-echelon setting, suppliers, intermediaries, and buyers continually reassess their relationships based on expected benefits and trust.
This integrated formulation provides a tractable foundation for analyzing how volatility and learning jointly shape persistence, reconfiguration, and stability in OSCs. We next describe each component in detail.

A summary of key symbols and notation used throughout the paper is provided in Table~\ref{tab:notations}.
\begin{table}[htbp]
\centering
\scriptsize
\caption{List of Notations}
\label{tab:notations}
\renewcommand{\arraystretch}{1.05}
\setlength{\tabcolsep}{4pt}
\begin{tabular}{p{2.2cm} p{11.3cm}}
\hline
\textbf{Symbol} & \textbf{Description} \\ \hline
$V$ & Set of agents (suppliers, intermediaries, buyers) \\
$V_t$ & Set of active agents at time $t$ \\
$E_t$ & Set of active contractual links at time $t$ \\
$G_t=(V_t,E_t)$ & Opportunistic supply chain (OSC) network at time $t$ \\
$p_i(t)$ & Stochastic unit price of supplier $i$ at time $t$ \\
$\mu$ & Drift term in Geometric Brownian Motion (GBM) \\
$\sigma$ & Volatility parameter in GBM \\
$W_i(t)$ & Standard Wiener process (Brownian motion) for supplier $i$ \\
$q_{ij}$ & Quantity procured by distributor $j$ from supplier $i$ \\
$q_i^{\max}$ & Maximum available supply quantity of supplier $i$ \\
$\beta_i(q)$ & Volume-based rebate or discount function \\
$c_i(q)=p_i(t)-\beta_i(q)$ & Effective procurement cost per unit \\
$\alpha,\,\beta$ & Demand function parameters ($P(q)=\alpha-\beta q$) \\
$P_{jk}(q)$ & Resale price function from distributor $j$ to buyer $k$ \\
$B_{jk}(t)$ & Belief (trust) of agent $j$ in downstream partner $k$ at time $t$ \\
$\lambda$ & Smoothing / learning-rate parameter in Bayesian updating \\
$\delta$ & Exogenous trust-decay rate in shock simulations \\
$\varphi_i^t$ & Product quality or reliability level of supplier $i$ at time $t$ \\
$\gamma$ & Perishability / quality decay factor \\
$E[\pi_j^t(i)]$ & Expected profit of agent $j$ from supplier $i$ at time $t$ \\
$\pi_j^t(i)$ & Realized profit of agent $j$ from supplier $i$ at time $t$ \\
$U_{ij}$ & Utility of maintaining link $(i,j)$ in the LOLOG model \\
$\pi_{ij}(\sigma;G)$ & Probability that link $(i,j)$ is active (link activation probability) \\
$F(\cdot)$ & Link-formation function in LOLOG (logistic transformation) \\
$\theta$ & Parameter vector in LOLOG capturing structural preferences \\
$(\theta_p,\theta_t,\theta_q)$ & LOLOG coefficients for profitability, trust, and structural dependence \\
$s(G)$ & Vector of network statistics (degree, reciprocity, clustering, etc.) \\
$I_i(G)$ & Influence index of node $i$ (trust-weighted profitability measure) \\
$S(\sigma)$ & Mean link-survival function as a function of volatility \\
$\sigma_c$ & Critical volatility threshold separating stable and unstable regimes \\
$MLSP(t)$ & Mean Link Survival Probability at time $t$ \\
$NCR(t)$ & Node Churn Rate at time $t$ \\
$\eta$ & Minimum positive transition probability in ergodicity proof \\
$\pi^*(G)$ & Stationary distribution of network configurations under LOLOG dynamics \\
$q^*$ & Optimal procurement quantity for a distributor \\
$\varepsilon$ & Minimum profitability threshold for maintaining a link \\
$\Delta_i$ & Perturbation vector in profit–trust space for node $i$ \\
$WTP_{jk}^t$ & Customer willingness-to-pay at link $(j,k)$ and time $t$ \\
$E_t^{jk}$ & Customer expenditure at time $t$ \\
$\pi_k^t(j)$ & Utility (or surplus) of customer $k$ buying from distributor $j$ at time $t$ \\
$N(G)$ & Candidate set of feasible links for activation or dissolution \\
$\Psi_{ij}(\sigma;G)$ & Economic (profit-based) component of link utility under volatility \\
$T_{ij}(\sigma;G)$ & Trust-based component of link utility under volatility \\
$Z_{ij}(G)$ & Structural component of link utility (degree, clustering, reciprocity) \\
$T$ & Simulation horizon (number of periods) \\
\hline
\end{tabular}
\end{table}

\subsection{Network Representation}

The OSC is represented as a directed, multi-echelon network observed at discrete time points 
$t \in \{0,1,2,\dots\}$. Let $V$ denote the set of agents (suppliers, intermediaries, and buyers), and let 
$E_t \subseteq V \times V$ denote the set of active directed links at time $t$. Each link $(i,j) \in E_t$ represents a 
contractual relationship through which goods, information, or payments flow from agent $i$ to agent $j$.
Agents may maintain multiple upstream and downstream relationships, with link formation restricted primarily to adjacent 
echelons to preserve the hierarchical structure of SCs. The network evolves dynamically as links are formed or 
dissolved in response to changes in expected profitability and adaptive trust suited to  opportunistic, high-churn supply environments explained in later subsections.

\subsection{Stochastic Pricing of Contracts: Geometric Brownian Motion (GBM)}\label{subsec_stoch_price}

In the absence of long-term contracts, prices in OSCs fluctuate continuously due to market volatility, demand shocks, and speculative trading. To represent this uncertainty, the unit price of an upstream agent \( i \) is modeled as a \emph{Geometric Brownian Motion (GBM)}:

\begin{equation}\label{eqn_GBM}
    dp_i(t) = \mu p_i(t)\,dt + \sigma p_i(t)\,dW_i(t),
\end{equation}
where \( \mu \) denotes the average drift (systematic trend), \( \sigma > 0 \) captures market volatility, and \( W_i(t) \) is a standard Wiener process. The analytical solution to \eqref{eqn_GBM} is given by (see \cite{oksendal2003}):

\begin{equation}\label{eqn_GBM_sol}
    p_i(t) = p_i(0)\exp\!\left[\!\left(\mu - \frac{\sigma^2}{2}\right)t + \sigma W_i(t)\!\right].
\end{equation}

GBM provides a parsimonious representation of short-horizon price dynamics relevant to OSCs. First, it guarantees strictly positive prices. Second, it captures the skewed volatility patterns observed in spot procurement markets. Third, it is consistent with environments where firms react to realized price shocks rather than smoothing prices toward a long-run equilibrium. These features align with high-churn, contract-free sourcing contexts such as fast fashion, electronics spot markets, and perishables.
While alternative stochastic processes (e.g., mean-reverting or jump-diffusion models) may be appropriate in settings with long-term contracts or infrequent extreme shocks, GBM serves as a tractable benchmark for analyzing decentralized decision-making under continuous volatility \cite{schwartz1997}. The proposed framework is  flexible to accommodate alternative price processes in future empirical extensions.

\subsection{Contract Structure and Expected Profit Computation}\label{subsec_contracts}

In OSCs, upstream agents offer short-term contracts that are frequently revised in response to market volatility, production capacity, and perceived reliability. Each supplier \( i \) at time \( t \) proposes a contract 
\[
    C_i^{t} = \left( p_i^{t}, q_i^{\max}, \phi_i^{t}, \beta_i(q) \right),
\]
where \( p_i^{t} \) denotes the stochastic unit price determined by the GBM process in~\eqref{eqn_GBM}, and \( q_i^{\max} \) represents the supplier’s maximum available quantity, reflecting production limits or inventory constraints. The term \( \phi_i^{t} \) captures product quality or reliability, shaped by technology and input variation, such that higher values of \( \phi_i^{t} \) enhance buyer preference and support premium pricing. The function \( \beta_i(q) \) denotes a volume-based rebate or discount that lowers the effective unit cost with larger order quantities, capturing standard scale economies observed in spot and short-term transactions. 

\textit{Distributor's Profit:}
Upon receiving a contract \( C_i^{t} \), downstream agent \( j \) (if it decides to accept the contract) determines an optimal procurement quantity \( q_{ij} \leq q_i^{\max} \), balancing procurement costs against uncertain resale revenues. The expected profit for agent \( j \) procuring from supplier \( i \) is expressed as:
\begin{eqnarray}\label{eqn_profit}
    \mathbb{E}\big[\pi_{j}^{t}(i)\big] 
    = 
    \sum_{k \in K_j} 
    \mathbb{E}\!\left[
        B_{jk}(t) \, P_{jk}(q_{jk}) \, q_{jk}
    \right]
      -
      \big(p_i^{t} - \beta_i(q_{ij})\big)\, q_{ij},
\end{eqnarray}
where \( K_j \) denotes the set of downstream customers of distributor \( j \) 
Here, \( P_{jk}(q_{jk}) \) represents the resale price offered to customer \( k \), 
which may depend on the traded volume or prevailing market conditions.

\vspace{0.2cm}
\textit{Economic interpretation of trust:}
The belief variable \( B_{jk}(t) \in (0,1) \) represents buyer \( j \)’s subjective probability that a downstream transaction with customer \( k \) is successfully fulfilled. Under incomplete contracting and weak enforcement, expected resale revenue equals the economic payoff \( P_{jk}(q_{jk}) q_{jk} \) weighted by this success probability. This multiplicative structure aligns with Bayesian interpretations of trust under uncertainty and with empirical evidence from opportunistic exchange environments. Beliefs \( B_{jk}(t) \) evolve over time according to the Bayesian updating scheme in Subsection~\ref{sec_belief_update}.

\vspace{0.2cm}
\textit{Rebate structure and its role in procurement decisions:}
The term \( \beta_i(q_{ij}) \) denotes a volume-based rebate or discount, which reduces the effective procurement cost according to
\[
c_i(q_{ij}) = p_i^{t} - \beta_i(q_{ij}).
\]
As rebates reduce both average and marginal procurement cost, \( \beta_i(q) \) directly affects the distributor’s optimal procurement quantity \( q_{ij} \).  
Larger orders are incentivized when trust is high or price volatility is low, whereas heightened uncertainty or low beliefs discourage consolidation.  
Thus, the rebate function not only shifts effective procurement cost but also influences the interior solution of the optimization problem in \eqref{eqn_profit}, playing a central role in the interaction between volatility, trust, and purchasing behavior (see Theorem~\ref{thm:procurement_utility}).

\textit{Customer's Utility:}
At the final echelon, customer \( k \) purchasing from distributor \( j \) derives utility as the net surplus between total willingness-to-pay and actual expenditure. Following standard microeconomic and supply chain formulations (see \cite{varian1992}), this surplus-based utility representation captures the consumer’s valuation of the traded quantity and serves as a consistent measure of demand responsiveness. Assuming a linear demand function commonly considered in SC literature (see \cite{berry2021,raknerud2007}),
\begin{equation}\label{eq:demand_function}
    P(q) = \alpha - \beta q, \qquad \alpha > 0, \; \beta > 0,
\end{equation}
the optimal demand quantity corresponding to distributor price \( p_j^{t} \) is
\begin{equation}\label{eq:optimal_quantity}
    q_{jk}^{*} = \frac{\alpha - p_j^{t}}{\beta}.
\end{equation}
The total willingness-to-pay $(WTP)$ is obtained by integrating the demand function:
\begin{equation}\label{eq:wtp}
    WTP_{jk}^{t} = \int_{0}^{q_{jk}^{*}} P(q)\,dq = \alpha q_{jk}^{*} - \frac{\beta}{2}\big(q_{jk}^{*}\big)^{2}.
\end{equation}
The corresponding expenditure is \( E_{jk}^{t} = \big(p_j^{t} - \beta_j(q_{jk}^{*})\big) q_{jk}^{*} \), where \( \beta_j(\cdot) \) denotes the volume-based rebate or discount function offered by distributor \( j \). 
Hence, consumer utility is expressed as the net surplus:
\begin{equation}\label{eq:consumer_utility}
    \pi_{k}^{t}(j) = WTP_{jk}^{t} - E_{jk}^{t}
    = \alpha q_{jk}^{*} - \frac{\beta}{2}\big(q_{jk}^{*}\big)^{2}
    - \big(p_j^{t} - \beta_j(q_{jk}^{*})\big) q_{jk}^{*}.
\end{equation}
Simplifying \eqref{eq:consumer_utility} under the assumption of a linear inverse demand 
function in \ref{eq:demand_function} yields
\begin{equation}\label{eq:utility_simplified}
    \pi_{k}^{t}(j) = (\alpha - p_j^{t}) q_{jk}^{*}
    - \frac{\beta}{2}\big(q_{jk}^{*}\big)^{2}
    + \beta_j(q_{jk}^{*}) q_{jk}^{*}.
\end{equation}
This expression holds specifically for the linear-demand setting and highlights how price 
sensitivity (\(\beta\)) and the rebate structure (\(\beta_j\)) together shape customer surplus 
and downstream demand responsiveness. The formulation thus provides a microeconomic link 
between consumer behavior, contract profitability, and network stability within the OSC 
framework.

\subsection{Bayesian Belief Updating in Opportunistic Supply Chains}
\label{sec_belief_update}

In OSCs, transactions are short-term and contractual enforcement is limited, requiring agents to continuously assess the reliability of their trading partners based on observed performance. Let $B_{jk}(t) \in [0,1]$ denote agent $j$'s belief at time $t$ regarding the reliability of partner $k$, and let $s_{jk}(t{+}1) \in \{0,1\}$ denote the observed outcome of their interaction, where $1$ indicates successful fulfillment.

Beliefs evolve according to the smoothed update
\begin{equation}\label{eq:smoothing_update}
B_{jk}(t{+}1)
= (1-\lambda)\,B_{jk}(t) + \lambda\,s_{jk}(t{+}1),
\end{equation}
where $\lambda \in (0,1)$ is a learning-rate parameter. This update provides a computationally efficient approximation to Bayesian learning under bounded rationality; full Bayesian derivations and alternative representations are provided in Appendix A.

\subsection{Network Evolution via Latent-Order-Logistic (LOLOG) Model}
\label{subsec_lolog}
Link formation and dissolution in the opportunistic supply chain are modeled using a latent-order logistic (LOLOG) formulation. Let $G_t$ denote the directed network at time $t$. The probability that a potential link $(i,j)$ is active, conditional on the current configuration, is
\begin{equation}\label{eqn_LOLOG}
P(G_t)
=
\frac{\exp\!\big(\boldsymbol{\theta}^\top \mathbf{s}(G_t + (i,j))\big)}
{\sum_{G'} \exp\!\big(\boldsymbol{\theta}^\top \mathbf{s}(G')\big)},
\end{equation}
where $\mathbf{s}(G) \in \mathbb{R}^d$ is a vector of network statistics and $\boldsymbol{\theta} \in \mathbb{R}^d$ captures agents’ sensitivities.
The statistic vector $\mathbf{s}(G_t)$ aggregates economic incentives, adaptive trust, and structural effects. Expected profits $E[\pi_j^t(i)]$ capture dyadic economic payoffs, trust beliefs reflect perceived partner reliability, and structural statistics such as reciprocity, degree, and clustering encode endogenous network dependence.
This formulation captures decentralized, stochastic link decisions driven by local profitability, trust, and network structure in OSCs. 

\subsubsection*{Comparison of LOLOG with Alternative Network Models}

Exponential Random Graph Models (ERGMs) and related equilibrium-based frameworks typically assume globally consistent configurations and simultaneous determination of network structure. These assumptions align with static or slowly evolving networks but are less appropriate for OSCs, where partner switching is rapid and decisions are locally driven. Dynamic bilateral matching and coalition formation models rely on stable preferences, transferable utilities, or forward-looking equilibrium behavior. Such requirements are difficult to be suitable for environments characterized by high volatility, short contracting horizons, and incomplete information. Similarly, multi-agent Markov decision processes and stochastic control formulations often assume system-wide observability and impose equilibrium or optimal policy computation under full rationality. These conditions become computationally prohibitive in large, evolving networks and are inconsistent with bounded rationality and decentralized decision-making in OSCs. In contrast, the LOLOG framework offers a probabilistic approach in which link updates depend only on local network structure, expected profitability, and adaptive trust. This enables tractable modeling of decentralized, high-churn network evolution while remaining consistent with stochastic pricing and learning dynamics suitable for OSCs.

\subsection{Remarks on the Integrated GBM-Bayesian-LOLOG Framework}
This integrated formulation links stochastic pricing, adaptive trust, and endogenous network restructuring within a unified behavioral process; price shocks influence profitability, profitability shapes trust, and evolving beliefs drive partner switching and structural adaptation, while network configuration feeds back into future volatility exposure. The model remains analytically tractable, reflects decentralized short-horizon decision making, and captures the high churn typical of OSCs. Practically, it provides a structured way to understand when networks remain conditionally stable and when they fracture, offering insight into how volatility management, trust-building, and structural design jointly affect supply chain resilience.

\section{Analysis}
\label{sec:analysis}

 This section presents a formal analysis of network stability, ergodicity, procurement behavior, and comparative statics within the integrated GBM--Bayesian--LOLOG framework developed in Section~\ref{sec_model}. All the proofs of the Theorems and Lemmas are in Appendix B. In what follows, stability is not interpreted as point-wise convergence of the network process $\{ G_t \}$ 
 to a fixed configuration. Rather, stability refers to the existence and persistence of statistically stable regimes, characterized by invariant or slowly varying distributions of network configurations and link-survival probabilities under stochastic evolution. This interpretation is consistent with the ergodic analysis in Subsection \ref{subsec_ergodicity} and the resilience metrics employed in Section \ref{sec:simulations}.

\subsection*{Assumptions}

We impose the following assumptions throughout the analysis:

\begin{description}
    \item[\textbf{A.1 (Finiteness).}] The agent set \( V \) is finite (i.e. $|V| < \infty$). Hence, the feasible set of network configurations 
    \( \mathcal{G} \subseteq \{0,1\}^{|V| \times |V|} \) is finite, ensuring a finite state space suitable for fixed-point and Markov chain analysis.

    \item[\textbf{A.2 (Price and revenue regularity).}] For each downstream pair \( (j,k) \), the resale price function \( P_{jk}(q) \) is continuously differentiable, strictly decreasing (\( P'_{jk}(q) < 0 \)), and concave (\( P''_{jk}(q) \le 0 \)). Expectations under GBM price dynamics \eqref{eqn_GBM}--\eqref{eqn_GBM_sol} are finite.

  \item[\textbf{A.3 (Procurement cost regularity).}] 
For each supplier \( i \), the effective procurement cost function 
\( c_i(q) = p_i^t - \beta_i(q) \) (as defined in Eq.~\eqref{eqn_profit}) is continuously differentiable with respect to quantity \( q \). 
The marginal procurement cost is nondecreasing, i.e., \( c_i'(q) \ge 0 \), and the cost function is convex, i.e., \( c_i''(q) \ge 0 \). 
These conditions ensure well-behaved cost dynamics and guarantee the existence of a unique and stable optimum in the distributor’s procurement decision.

   \item[\textbf{A.4 (Interior solution).}] 
The feasible procurement quantity \( q \) lies within the closed interval \( [0, q^{\max}] \). 
Whenever the optimal solution is interior, i.e., \( 0 < q^\ast < q^{\max} \), it satisfies the necessary first-order optimality condition derived from the distributor’s expected utility maximization problem. 
This assumption ensures differentiability and tractability of the optimality analysis under standard regularity conditions.

\end{description}

\subsection{Existence of Stable Network Configurations}

Before analyzing long-run network properties, it is necessary to establish 
that the OSC admits at least one 
self-consistent configuration in which agents have no incentive to 
unilaterally alter their contracting relationships. 
This subsection states the additional assumptions specific to 
Theorem~\ref{thm:stable_network} and provides a formal proof of existence.

In addition to the general assumptions~\textbf{A.1--A.4} which are assumed throughout the paper, we have the following assumptions which are specific to the theorem we will state next. Suppose that:
\begin{description}
    \item [\textbf{S.1(Continuity of expected profits)}] 
    For each potential directed link \((i,j)\), 
    the expected distributor profit at time~\(t\), denoted 
    \(\mathbb{E}[\pi_j^t(i;G)]\), is continuous in the network configuration 
    \(G \in \mathcal{G}\), in the stochastic price process (modeled as 
    GBM), and in agents’ belief parameters that evolve 
    via Bayesian updating.
    
    \item [\textbf{S.2(Profitability threshold)}]  
    There exists a fixed constant \(\varepsilon > 0\) such that 
    any link \((i,j)\) with 
    \(\mathbb{E}[\pi_j^t(i;G)] \le \varepsilon\) 
    is not sustained (dissolved).
\end{description}

\begin{theorem}[Existence of stable network configurations]
\label{thm:stable_network}
Under Assumptions~\textbf{A.1--A.4} and~\textbf{S.1--S.2}, 
there exists at least one stable network configuration 
\(G^* \in \mathcal{G}\) such that:
\begin{enumerate}[(i)]
    \item \textbf{(Profitability)}  
    Every active link \((i,j) \in G^*\) satisfies 
    \(\mathbb{E}[\pi_j^t(i;G^*)] > \varepsilon\);
    \item \textbf{(Local stability)}  
    No agent \(i \in V\) can unilaterally modify its set of outgoing links in 
    \(G^*\) to achieve strictly higher expected profit, 
    subject to the profitability threshold~\(\varepsilon\).
\end{enumerate}
Consequently, the OSC admits at least one 
self-consistent configuration that remains stable even in the presence of 
stochastic pricing, decentralized decisions, and boundedly rational belief 
updates.
\end{theorem}
\textit{Remarks:}
Theorem~\ref{thm:stable_network} establishes the existence of at least one locally stable sub-network in an OSC under decentralized link decisions driven by profit and trust. Since the set of feasible network configurations $\mathcal{G}$ is finite, improvement paths generated by unilateral link additions or removals that strictly increase expected profit cannot cycle. Assumptions~ \textbf{S.1} and  \textbf{S.2} ensure well-defined profit changes under link toggles and rule out persistently unprofitable connections. Consequently, there exists at least one maximal configuration $G^\ast$ with no profitable unilateral deviation.
The stability identified is local rather than strategic: the result does not characterize Nash or pairwise equilibria, but instead identifies natural rest points of the GBM–Bayesian–LOLOG dynamics around which long-run behavior and resilience analysis can be organized.

\subsection{Critical Volatility Threshold for Network Stability}
\label{subsec_volatility_threshold}

While Theorem~\ref{thm:stable_network} guarantees the existence 
of at least one stable configuration, real-world OSCs often undergo 
phase-like transitions, moving from stable to unstable regimes as market
volatility increases. To characterize this transition formally, we define
the mean link-survival function \(S(\sigma)\) that summarizes the average
probability of maintaining contractual ties under different levels of price
volatility. We then identify a critical volatility threshold \(\sigma_c\)
beyond which opportunistic links dissolve faster than they can form,
causing the network to fragment.
We impose the following assumptions  governing volatility, link utilities,
and averaging behavior.

\subsection*{Definitions and Assumptions for Volatility-Dependent Link Survival}


\begin{definition}[Candidate link set]
For any feasible network configuration $G \in \mathcal{G}$, the set of
feasible candidate links is
\[
\mathcal{N}(G) \subseteq V \times V,
\]
consisting of all directed pairs $(i,j)$ that may form a contractual
relationship given the technological and echelon constraints of the supply
chain. By Assumption \textbf{A.1} (finite agent set), $\mathcal{N}(G)$ is finite and
nonempty for every~$G$.
\end{definition}

\begin{definition}[Link utility]
For any $(i,j)\in \mathcal{N}(G)$ and volatility level $\sigma\ge 0$, the
utility of maintaining or forming link $(i,j)$ is
\begin{equation}
U_{ij}(\sigma;G)
= \Psi_{ij}(\sigma;G) + T_{ij}(\sigma;G) + Z_{ij}(G),
\label{eqn_link_utility}
\end{equation}
where $\Psi_{ij}$, $T_{ij}$, and $Z_{ij}$ denote the economic, trust, and
structural components, respectively. This additive formulation separates the
distinct drivers of link choice in OSCs; profitability
under volatility, beliefs about partner reliability, and local network
structure and follows standard practice in LOLOG and ERGM models in which
utilities combine covariate effects and structural statistics linearly
\citep{Snijders2010,Fellows2022}. Empirical evidence similarly
shows that economic incentives, trust, and embedded structural positions
influence inter-organizational link choices independently and jointly
\citep{GulatiNickerson2008,SorensonWaguespack2006}.
\end{definition}


\begin{description}

\item[\textbf{B.1 (LOLOG activation rule).}]
For any $(i,j)\in \mathcal{N}(G)$, the activation probability is
\[
\pi_{ij}(\sigma;G) = F\!\big(U_{ij}(\sigma;G)\big),
\]
where $F:\mathbb{R}\to (0,1)$ is continuous and strictly increasing.  
This follows standard practice in LOLOG and dynamic ERGM formulations, where
$F$ typically corresponds to a logistic or probit link–response function
\citep{Snijders2010,Fellows2022}. Continuity ensures that small
changes in utility lead to smooth adjustments in activation probability, and
strict monotonicity guarantees that higher utility always increases the
likelihood of link formation. These properties are essential for the
continuity and monotonicity arguments underlying Theorem~\ref{thm:volatility}.

\item[\textbf{B.2 (Risk-adjusted utility).}]
The components of the decomposition in \eqref{eqn_link_utility} are
continuous in their arguments, and the economic term $\Psi_{ij}(\sigma;G)$ is
nonincreasing in $\sigma$. This captures the standard notion that greater
volatility reduces risk-adjusted profitability.

\item[\textbf{B.3 (Volatility-dependent trust).}]
The trust component $T_{ij}(\sigma;G)$ is continuous and nonincreasing in
$\sigma$, reflecting reduced perceived partner reliability under greater
volatility and ensuring that the composite mapping $F\circ U_{ij}$ remains
continuous.

\item[\textbf{B.4 (Finite averaging).}]
Because $\mathcal{N}(G)$ is finite for every $G\in \mathcal{G}$, the mean
link-survival function
\[
S(\sigma)
= \frac{1}{|\mathcal{G}|}\sum_{G\in\mathcal{G}}
  \frac{1}{|\mathcal{N}(G)|}\sum_{(i,j)\in\mathcal{N}(G)}
  \pi_{ij}(\sigma;G)
\]
is a finite sum of continuous functions and therefore continuous.

\end{description}

\begin{theorem}[Critical volatility threshold for network stability]
\label{thm:volatility}
 Under Assumptions~\textbf{A.1--A.4} and
\textbf{B.1--B.4}, define the mean link-survival function
\[
S(\sigma)
=\frac{1}{|\mathcal{G}|}\sum_{G\in\mathcal{G}}
  \frac{1}{|\mathcal{N}(G)|}
  \sum_{(i,j)\in\mathcal{N}(G)} \pi_{ij}(\sigma;G),
  \qquad \sigma\ge0.
\]
Then:
\begin{enumerate}[(i)]
\item \(S(\sigma)\) is continuous and nonincreasing on \([0,\infty)\);\vspace{1mm}
\item If \(S(0)>\displaystyle\lim_{\sigma\to\infty}S(\sigma)\), then for any
threshold \(s^*\) with
\(\displaystyle\lim_{\sigma\to\infty}S(\sigma)<s^*<S(0)\) there exists a
(finite) critical volatility level \(\sigma_c\ge0\) such that
\[
S(\sigma)\ge s^* \quad\text{for all } \sigma<\sigma_c,
\qquad
S(\sigma)< s^* \quad\text{for all } \sigma>\sigma_c,
\]
where 
\(\displaystyle \sigma_c := \inf\{\sigma\ge0:\ S(\sigma)<s^*\}.\)
\end{enumerate}
\end{theorem}
\qed

\textit{Remarks:}
Theorem~\ref{thm:volatility} shows that increasing volatility reduces expected profitability and trust, leading to a monotonic decline in the mean link-survival function $S(\sigma)$ under the LOLOG dynamics. Under the stated continuity and bounded-rationality assumptions, this implies the existence of a critical volatility level $\sigma_c$ beyond which link dissolution dominates link formation, signaling a transition from a stable to a fragmented network regime. The result does not imply a sharp mathematical phase transition, but rather identifies a tipping region that guides interpretation of OSC persistence under volatility and motivates the subsequent simulation and managerial analyses. A complete proof is provided in Appendix~B.

\subsection{Node Importance and Structural Influence}
\label{subsec:node_importance}

While the network-level stability results 
(Theorems~\ref{thm:stable_network} and~\ref{thm:volatility}) 
characterize aggregate persistence under stochastic volatility, 
it is also essential to understand how individual agents 
contribute to the overall robustness of the OSC.  
In practice, certain nodes typically those combining high profitability, reliability, 
and connectivity exert disproportionate influence on network stability.  
This subsection formalizes such effects through an \emph{influence index}, 
linking local trust-weighted profitability to global structural resilience.

\noindent\textbf{ C.1 (Differentiability and Monotonicity).}
For each active link \((i,j)\in E\),  
the LOLOG link probability \(\pi_{ij}(\sigma)\) introduced in 
Assumptions~\textbf{B.1--B.2} 
is continuously differentiable in its underlying arguments;
the expected profitability \(\mathbb{E}[\pi_j^t(i)]\) 
and the trust belief \(B_{ij}(t)\).  
It satisfies
\[
\frac{\partial \pi_{ij}}{\partial \mathbb{E}[\pi_j^t(i)]} > 0,
\qquad
\frac{\partial \pi_{ij}}{\partial B_{ij}(t)} > 0,
\]
implying that higher expected profits or stronger trust beliefs 
strictly increase the probability of maintaining link \((i,j)\).

\begin{theorem}[Node Importance in OSCs]
\label{thm:node_importance}
Consider an OSC network \(G=(V,E)\) governed by the 
GBM--Bayesian--LOLOG framework.  
For each node \(i\in V\), define its expected \emph{influence index} as
\[
I_i(G)
=\sum_{j:(i,j)\in E}\mathbb{E}[\pi_j^t(i)]\,B_{ij}(t),
\]
representing the trust-weighted expected contribution of node \(i\) 
to its downstream partners.  
Under Assumptions~\textbf{A.2--A.3} and \textbf{C.1}, the following statements hold:
\begin{enumerate}
    \item The influence index \(I_i(G)\) is continuous and nondecreasing 
    in both expected profitability and downstream trust;
    \item Node \(i\) is \emph{structurally critical} if an increase in its expected profit or trust 
    raises the network stability function \(S(\sigma)\) (as defined in Theorem~\ref{thm:volatility}), i.e.,
    \[
    \frac{\partial S(\sigma)}{\partial I_i(G)} > 0,
    \]
    implying that node \(i\) exerts a positive marginal effect on long-run network persistence.
\end{enumerate}
\end{theorem}
\qed

\textit{Remarks:}
Theorem~\ref{thm:node_importance} follows from the fact that link persistence in the LOLOG model increases with expected profitability and trust. Since an agent’s influence index $I_i(G)$ aggregates its trust-weighted expected contributions to downstream partners, nodes with higher $I_i(G)$ naturally support stronger link retention and therefore enhance the overall network stability. When such influential nodes weaken or exit, the probability of link continuation decreases along their connected neighborhoods, reducing the aggregate survival measure $S(\sigma)$ and making fragmentation more likely. Thus, the theorem characterizes how micro-level agent performance translates into macro-level resilience. This result is useful in our framework as it highlights which agents function as structural anchors and explains why network breakdown often originates around the loss of trusted, profitable hubs.

\subsection{Profit Sensitivity to Trust Beliefs}
Trust plays a central role in OSC coordination because contracts are short-term and enforcement is limited.
Beliefs $B_{jk}$
 represent the distributor’s perception of downstream reliability, which evolves adaptively through Bayesian updating.
This subsection examines the local effect of trust on expected profit while holding procurement quantities fixed.
We show that expected profit is non-decreasing in trust beliefs, thereby quantifying the direct informational benefit of confidence in downstream partners.

\begin{lemma}[Profit monotonicity in beliefs]
\label{lem:beliefs_updated}
Assume \textbf{A.2} (price and revenue regularity) and consider the distributor’s expected profit function in Eq.~\eqref{eqn_profit}.  
Fix procurement quantities \(\{q_{jk}\}_k\) and \(q_{ij}\).  
Then for any downstream index \(\kappa\),
\[
\frac{\partial}{\partial B_{j\kappa}(t)}\,\mathbb{E}\!\big[\pi_j^{t}(i)\big]
= \mathbb{E}\!\big[P_{j\kappa}(q_{j\kappa})\,q_{j\kappa}\big].
\]
Consequently \(\mathbb{E}[\pi_j^{t}(i)]\) is nondecreasing in each belief \(B_{jk}(t)\),
and it is strictly increasing in \(B_{j\kappa}(t)\) whenever \(\mathbb{E}[P_{j\kappa}(q_{j\kappa})\,q_{j\kappa}]>0\).
\end{lemma}
\qed

 \textbf{ Proof of Lemma \ref{lem:beliefs_updated}: }
From~\eqref{eqn_profit} the distributor's expected profit at time \(t\) may be written as
\[
\mathbb{E}[\pi_j^t(i)]
= \mathbb{E}\Big[ \sum_k B_{jk}(t)\,P_{jk}(q_{jk})\,q_{jk}
      - \big(p_i^t-\beta_i(q_{ij})\big)q_{ij}\Big].
\]
As \(B_{jk}(t)\) is \(\mathcal{F}_t\)-measurable (it is the agent's posterior at decision time), we may take conditional expectations with respect to \(\mathcal{F}_t\) and treat \(B_{jk}(t)\) as a deterministic coefficient inside the conditional expectation. Hence
\[
\mathbb{E}[\pi_j^t(i)]
= \sum_k B_{jk}(t)\,\mathbb{E}\!\big[P_{jk}(q_{jk})\,q_{jk}\big]
  - \big(p_i^t-\beta_i(q_{ij})\big)q_{ij},
\]
where expectations are unconditional (or equivalently conditional on \(\mathcal{F}_t\)) by linearity.

Assumption~\textbf{A.2} provides the integrability and differentiability conditions needed to apply dominated convergence, so differentiation with respect to \(B_{j\kappa}(t)\) may be interchanged with expectation. Differentiating the above expression gives
\[
\frac{\partial}{\partial B_{j\kappa}(t)}\mathbb{E}[\pi_j^t(i)]
= \mathbb{E}\!\big[P_{j\kappa}(q_{j\kappa})\,q_{j\kappa}\big].
\]
Since \(P_{j\kappa}(q_{j\kappa})\) and \(q_{j\kappa}\) are nonnegative by model construction, the right-hand side is nonnegative, establishing monotonicity. Moreover, if \(\mathbb{E}[P_{j\kappa}(q_{j\kappa})\,q_{j\kappa}]>0\), the derivative is strictly positive, giving strict increase in that direction.
\qed

\textit{Remarks:}
Lemma~\ref{lem:beliefs_updated} shows that trust affects profits through an informational channel.  
When a distributor has stronger beliefs about the reliability of downstream partners, expected revenues increase even before any adjustment in production or procurement decisions.

\subsection{Optimal Procurement Behavior}
Having established the effect of beliefs on expected profit, we now turn to the distributor’s procurement decision, the key behavioral choice driving OSC dynamics.
Here, the distributor optimizes expected utility in \eqref{eqn_profit}
, trading off stochastic resale revenues and effective procurement costs that depend on supplier price volatility and volume-based rebates.
We analyze two core properties: (i) the existence and uniqueness of an optimal procurement quantity, and (ii) the monotone comparative statics of this optimum with respect to beliefs and supplier prices.
Together, these results characterize economically consistent and stable sourcing behavior in uncertain environments.

\begin{theorem}[Existence and uniqueness of optimal procurement]
\label{thm:procurement_utility}
Under Assumptions~\textbf{A.2--A.4}, 
the distributor’s problem
\[
\max_{q\in[0,q^{\max}]}\;
\mathcal{U}_j(q)
=\sum_{k\in\mathcal{K}_j}\mathbb{E}[B_{jk}P_{jk}(q)q]
- c_i(q)q,
\qquad 
c_i(q)=p_i^t-\beta_i(q),
\]
admits an unique optimal solution \(q^\ast\in[0,q^{\max}]\).  
If \(0<q^\ast<q^{\max}\), it satisfies the first-order condition
\[
\sum_{k\in\mathcal{K}_j}
\frac{\partial}{\partial q}\,
\mathbb{E}[B_{jk}P_{jk}(q)q]\Big|_{q=q^\ast}
=c_i'(q^\ast)q^\ast+c_i(q^\ast).
\]
\end{theorem}
\qed

\paragraph{Remarks:}
Theorem~\ref{thm:procurement_utility} ensures well-posed distributor behavior in OSCs: given beliefs and prices, procurement decisions are unique and economically interpretable as balancing expected marginal revenue against marginal procurement cost.

\begin{theorem}[Monotone comparative statics of the optimal procurement quantity]
\label{thm:comparative}
Under Assumptions~\textbf{A.2--A.4}, 
let \(q^\ast(B,p_i^t)\) denote the unique interior maximizer of the distributor’s problem~\eqref{eqn_profit}.  
Then:
\begin{enumerate}[(i)]
    \item The optimal procurement quantity \(q^\ast\) is nondecreasing in each belief parameter \(B_{jk}\);\footnote{%
    The monotonicity with respect to \(B_{jk}\) holds under the mild local condition 
    \(\tfrac{d}{dq}\mathbb{E}[P_{jk}(q)q]\big|_{q=q^\ast}\ge0\), 
    which ensures that increased trust enhances expected marginal revenue at the optimum.}
    \item \(q^\ast\) is nonincreasing in the supplier price \(p_i^t\).
\end{enumerate}
\end{theorem}
\qed

\textit{Remarks:}
Theorem~\ref{thm:comparative} formalizes intuitive behavioral responses of the distributor’s optimal procurement decisions. 
An increase in the belief parameter $B_{jk}$ reflecting greater trust in downstream partners raises expected marginal revenue and thus increases the optimal procurement quantity $q^\ast$. 
Conversely, a higher supplier price $p_i^t$ elevates marginal cost, leading to a lower $q^\ast$. 
These monotone adjustments capture the adaptive coordination logic of opportunistic supply chains, wherein agents revise procurement decisions in response to evolving trust and price signals to sustain efficiency under uncertainty.

\subsection{Long-Run Network Behavior: Ergodicity}\label{subsec_ergodicity}
 While stability and procurement results describe local equilibrium behavior, the OSC network evolves stochastically over time.
This subsection analyzes its ergodic properties—that is, whether the dynamic network converges to a stationary distribution independent of initial conditions.
Using the LOLOG process and mild positivity assumptions on transition probabilities, we show that the resulting Markov chain is finite, irreducible, and aperiodic, thus admitting a unique stationary distribution.
This guarantees that empirical or simulated network statistics converge to their
true long-run values, enabling reliable steady-state analysis of OSC structures.
In practical terms, it means that observed outcomes in simulation are not artifacts
of initial conditions, but reflect the inherent long-run behavior of the system; 
a key requirement for drawing meaningful resilience and policy insights.
\begin{theorem}[Ergodicity of LOLOG Network Dynamics]
\label{thm:ergodicity}
Let $\{G_t\}_{t\ge 0}$ denote the sequence of OSC network configurations
generated by the LOLOG update rule~\eqref{eqn_LOLOG}, where each 
$G_t\in\mathcal{G}$, the finite set of all feasible network configurations.
Assume \textbf{A.1} (finiteness of the agent set) and suppose that there 
exists a constant $\eta>0$ such that for every pair of configurations 
$G,G'\in\mathcal{G}$:
\begin{enumerate}
  \item any feasible single-edge toggle transforming $G$ into $G'$ 
        (addition or deletion of one link) occurs with probability 
        at least $\eta$;
  \item the self-transition probability satisfies $P(G\to G)\ge \eta$ \footnote{The uniform positivity constant $\eta>0$ is used for
 modeling small random perturbations (e.g., exploration noise or a
trembling-hand) embedded in the LOLOG updating mechanism; it ensures that
every feasible single-edge toggle remains possible with nonzero probability,
which yields irreducibility and aperiodicity of the induced Markov chain.}
\end{enumerate}
Then the process $\{G_t\}$ forms a finite-state, irreducible, and aperiodic 
Markov chain. Hence it admits a unique stationary distribution $\pi^\ast$ 
on $\mathcal{G}$, and for any bounded function $f:\mathcal{G}\to\mathbb{R}$,
\begin{eqnarray}\label{eq:ergodic_convergence}
    \frac{1}{T}\sum_{t=1}^{T} f(G_t)
\;\xrightarrow{\mathrm{a.s.}}\;
\sum_{G\in\mathcal{G}} f(G)\,\pi^\ast(G),
\qquad T\to\infty.
\end{eqnarray}
\end{theorem}

\textit{Remarks:}
Theorem~\ref{thm:ergodicity} establishes that the LOLOG network dynamics form an ergodic finite-state Markov chain over OSC configurations, implying convergence to a unique stationary distribution $\pi^\ast$. While individual links may continue to form and dissolve, the distribution of network structures stabilizes over time. Consequently, long-run averages of bounded network statistics such as degree, clustering, or link persistence—converge almost surely to their expectations under $\pi^\ast$, allowing resilience measures computed from long simulations to be interpreted as steady-state properties rather than transient artifacts.

\section{Numerical Experiments}
\label{sec:simulations}

This section presents the numerical implementation and simulation-based evaluation of the integrated GBM-Bayesian-LOLOG framework developed in Section~\ref{sec_model}. The objective is to examine how OSCs evolve under stochastic prices, adaptive trust, and decentralized link formation, and to assess whether the qualitative stability mechanisms predicted in Section~\ref{sec:analysis} emerge under realistic, industry-informed conditions.

The numerical experiments employ industry-informed parameter ranges designed to reflect stylized characteristics of OSCs . Exploratory discussions with supply chain managers in fast fashion, electronics, and perishable sectors were used only to identify plausible magnitudes of volatility, contracting practices, trust dynamics, and sourcing structures; they were not used for calibration or statistical estimation. The simulations should therefore be interpreted as computational experiments illustrating qualitative implications of the framework, rather than empirical validation. All parameter values are taken directly from Tables~\ref{tab:global_params} and~\ref{tab:industry_params}, ensuring internal consistency while capturing differences in volatility, governance, and perishability.

To provide a broad basis for comparison, we examine three industries representing distinct OSC environments: \textit{fast-fashion apparel}, \textit{electronics component spot markets}, and \textit{perishable agricultural goods}. Each industry exhibits a characteristic combination of price volatility, relational governance, and product attributes. The fast-fashion setting is informed by on-site discussions in the Surat textile cluster, a globally significant hub for rapid-cycle fabric and garment sourcing, where opportunistic switching is common but embedded within a persistent relational core. The electronics sector reflects the volatility amplification and supplier concentration observed during the 2020--2022 global semiconductor shortage \cite{semiconductor_shortage,sia_supply_chain}, while the perishable agricultural case captures decay-driven fragility documented in agri-food supply systems \cite{oecd_agri_volatility,fao_food_volatility}.

All scenarios employ a fixed tripartite network of 5 suppliers, 8 distributors, and 14 consumers, simulated over $T=100$ periods. Two environments are considered: a \emph{baseline} case and a \emph{shock scenario} incorporating exogenous price spikes, trust collapses, and firm exits. The shock scenario stresses the system to test whether stability mechanisms persist under compound disturbances. In both environments, prices, beliefs, and network structure co-evolve endogenously through repeated LOLOG-based partner evaluation.
\subsection{Resilience Metrics}
\label{subsec:resilience_metrics}

To connect simulation outcomes with the stability conditions in Section~\ref{sec:analysis}, we employ two complementary metrics: the \textit{Mean Link Survival Probability} (MLSP) and the \textit{Node Churn Rate} (NCR). MLSP measures relationship persistence by quantifying the proportion of buyer--supplier links that survive from period to period, directly corresponding to the survival function $S(\sigma)$ appearing in Theorem~\ref{thm:ergodicity}. NCR measures volatility in agent participation by tracking entries and exits in each layer of the supply chain.

Formally, if $E_t$ and $V_t$ denote the link set and node set at time $t$,
\[
\text{MLSP}(t)=
\begin{cases}
\frac{|E_{t-1}\cap E_t|}{|E_{t-1}|}, & |E_{t-1}|>0,\\
0, & |E_{t-1}|=0,
\end{cases}
\qquad
\text{NCR}(t)=\frac{|V_{t-1}\triangle V_t|}{|V_{t-1}|}.
\]
Their temporal averages
\[
\overline{\text{MLSP}}=\frac{1}{T}\sum_{t=1}^T \text{MLSP}(t),
\qquad
\overline{\text{NCR}}=\frac{1}{T}\sum_{t=1}^T \text{NCR}(t)
\]
provide summary measures of structural persistence and actor-level turbulence. As implied by Theorem~\ref{thm:ergodicity}, $\overline{\text{MLSP}}$ empirically estimates the long-run survival probability of relationships under industry-specific volatility.

\subsection{Simulation Inputs and Calibration}
\label{subsec:inputs}

Tables~\ref{tab:global_params} and \ref{tab:industry_params} list the complete parameterization used in all runs. At the onset of this study, we conducted in-depth discussions with senior managers across  sectors such as fast fashion and semiconductor industry to understand volatility patterns, contracting practices, trust dynamics, quality decay, and typical sourcing structures observed in practice. All simulations directly employ the parameter values listed in Tables~\ref{tab:global_params} and \ref{tab:industry_params}, including the perishable decay factor ($\gamma = 0.90$), shock probability ($0.2$), trust decay under shocks ($0.85$), and the industry-specific LOLOG coefficients $(\theta_p,\theta_t,\theta_q)$. These parameter values are directly informed by (i) field observations in Surat (fast fashion) and also industry reports \cite{mckinsey_fashion_2025}, (ii) industry reports on semiconductor volatility \cite{semiconductor_shortage,sia_supply_chain}, and (iii) perishability studies from FAO and OECD \cite{oecd_agri_volatility,fao_food_volatility}. The fixed network size ensures consistent comparability across experiments, while the longer horizon ($T=100$) enables clear identification of asymptotic trends predicted by Theorem~\ref{thm:volatility}.
\begin{table}[H]
\centering
\resizebox{0.85\textwidth}{!}{
\begin{tabular}{ll}
\toprule
Parameter & Value / Description \\ 
\midrule
Network size & 5 suppliers, 8 distributors, 14 consumers \\
Simulation horizon & $T=100$ periods \\
Price process & GBM with industry-specific $(\mu,\sigma)$ \\ 
Trust updating & Bayesian smoothing with type-specific learning rate $\lambda$ \\
LOLOG coefficients & $(\theta_p,\theta_t,\theta_q)$ from Table~\ref{tab:industry_params} \\
Shock probability & 0.2 per period \\
Shock types & Price spike, node exit, trust collapse \\
Trust decay in shocks & $0.85$ \\
Perishability factor & $\gamma=0.90$ (agriculture only) \\
Supplier capacity & $q_{\max}\in[8,15]$ \\
Rebate function & $\beta_i(q)=\delta_i\sqrt{q}$ (concave volume rebate) \\
\bottomrule
\end{tabular}}
\caption{Global simulation parameters.}
\label{tab:global_params}
\end{table}

\begin{table}[H]
\centering
\resizebox{0.8\textwidth}{!}{
\begin{tabular}{lccccccc}
\toprule
Industry & $\sigma$ & $\mu$ & Baseline Trust & Agent-Type Distribution &
LOLOG $(\theta_p,\theta_t,\theta_q)$ & Perishability \\ 
\midrule
Fast Fashion & 0.30 & 0.02 & 0.55 &
$[0.50,0.20,0.10,0.20]$ & $(2.0,0.3,0.15)$ & No \\
Electronics & 0.70 & 0.05 & 0.35 &
$[0.60,0.10,0.05,0.25]$ & $(3.5,0.1,0.40)$ & No \\
Perishables & 0.40 & 0.01 & 0.75 &
$[0.40,0.25,0.25,0.10]$ & $(2.5,0.4,0.80)$ & Yes ($\gamma=0.90$) \\
\bottomrule
\end{tabular}}
\caption{Industry-specific calibration values (Table~3).}
\label{tab:industry_params}
\end{table}

\subsection{Fast-Fashion Apparel}
\label{subsec:fastfashion}

\begin{figure}[H]
\centering
\includegraphics[width=0.48\textwidth]{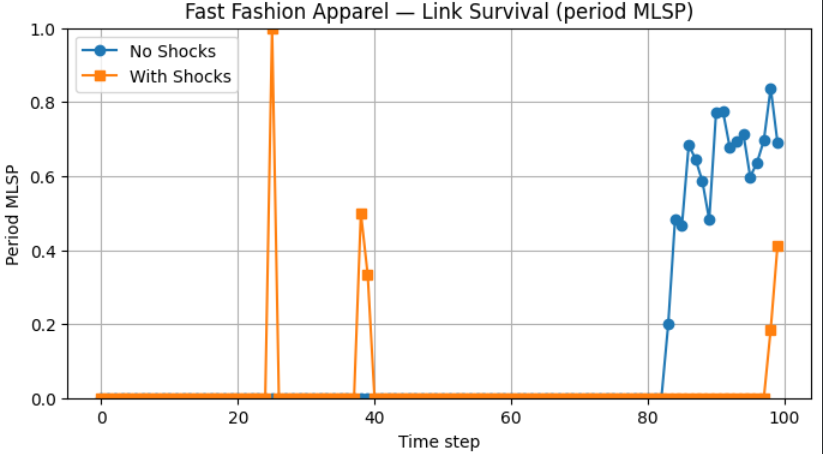}
\includegraphics[width=0.48\textwidth]{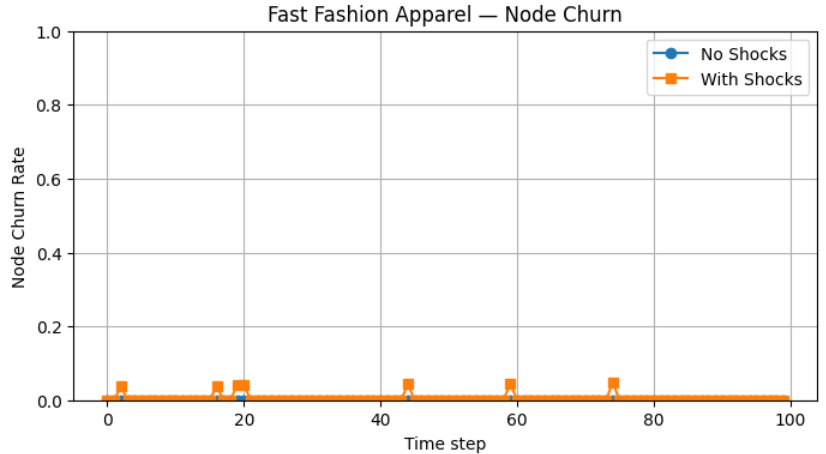}
\caption{Fast-fashion apparel: MLSP and NCR across 100 periods with and without shocks.}
\label{fig:fashion_mslp}
\label{fig:fashion_ncr}
\end{figure}

\begin{figure}[H]
\centering
\includegraphics[width=0.48\textwidth]{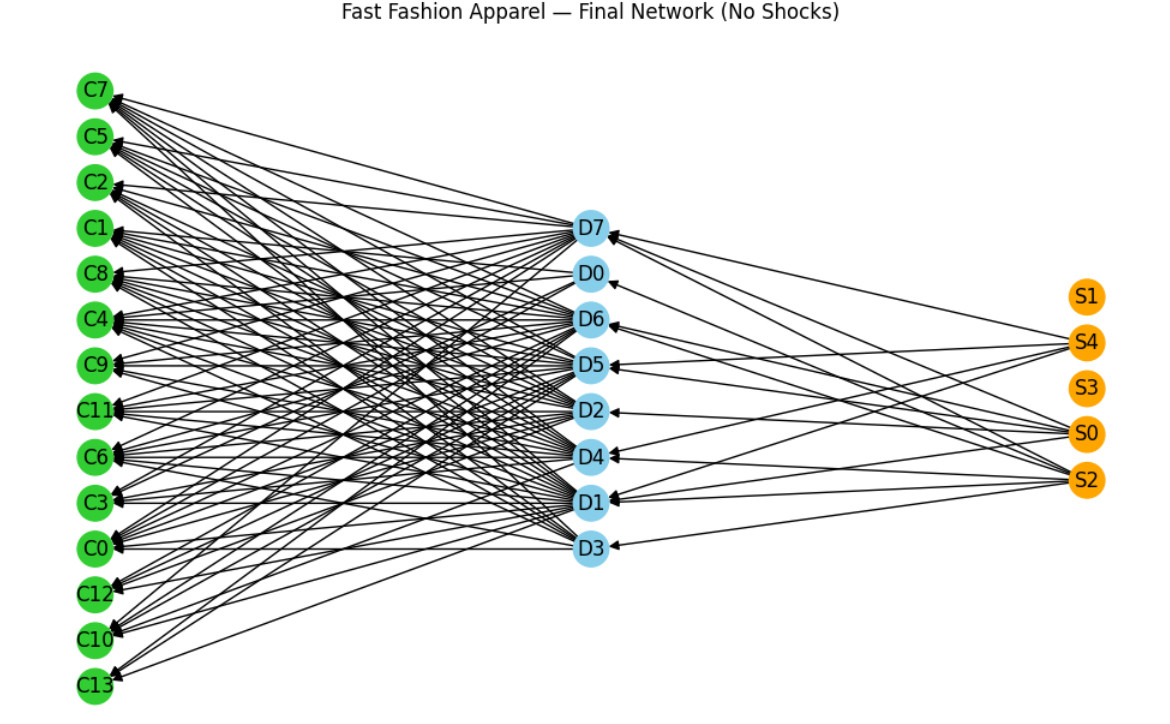}
\includegraphics[width=0.48\textwidth]{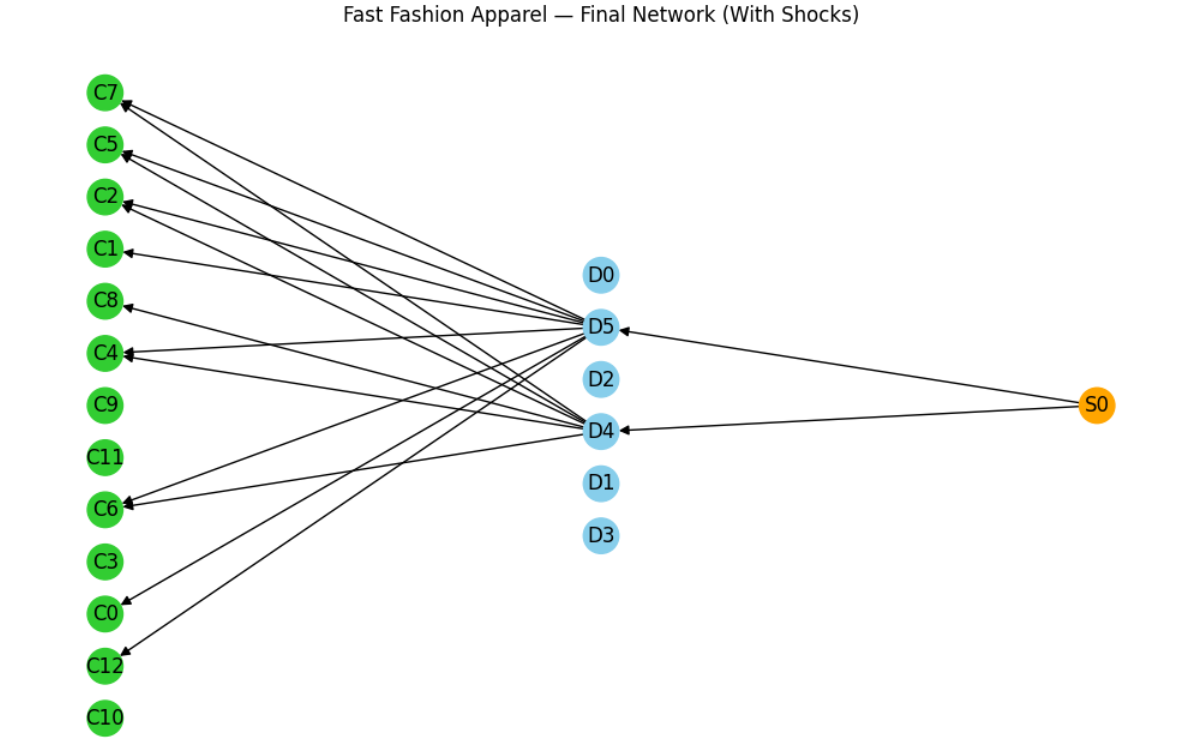}
\caption{Fast-fashion apparel: final network structures under both scenarios.}
\label{fig:fashion_networks}
\end{figure}

The fast-fashion network exhibits the highest degree of adaptive resilience. As seen in Figure~\ref{fig:fashion_mslp}, the MLSP remains relatively stable between the baseline ($\approx 0.75$) and shock scenario ($\approx 0.60$). NCR stays extremely low (Figure~\ref{fig:fashion_ncr}), reflecting that although links frequently rewire due to opportunistic sourcing, firms themselves rarely exit. This behavior is visually evident in the final networks (Figure~\ref{fig:fashion_networks}), where the structure remains dense under both conditions.
These patterns closely match the empirical observations from the Surat textile cluster, where firms maintain a small pool of trusted long-term partners while opportunistically adjusting transactional ties based on weekly price shifts. Theoretical predictions are thus validated: with moderate volatility ($\sigma=0.30<\sigma_c$), the network achieves the stable yet adaptive regime characterized by Theorem~\ref{thm:stable_network}.

\subsection{Electronics Component Spot Markets}
\label{subsec:electronics}

\begin{figure}[H]
\centering
\includegraphics[width=0.48\textwidth]{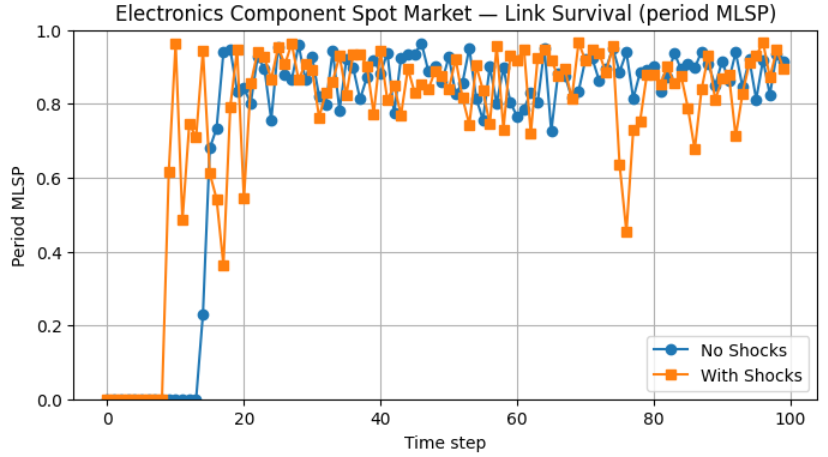}
\includegraphics[width=0.48\textwidth]{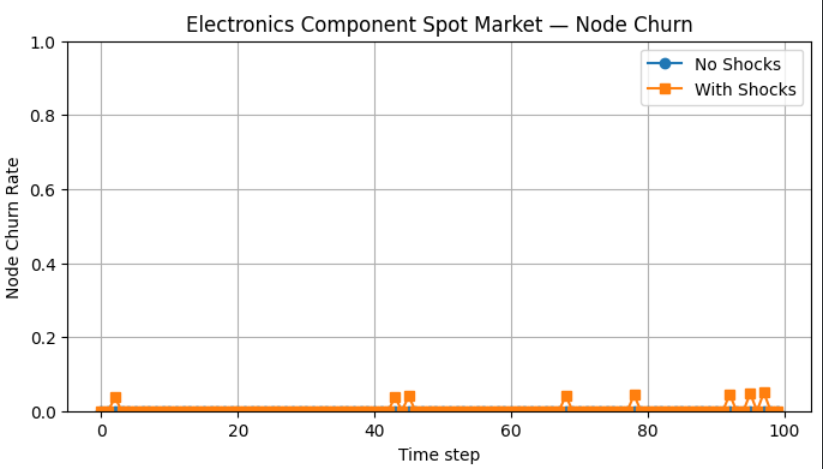}
\caption{Electronics spot market: MLSP and NCR across 100 periods.}
\label{fig:elec_mslp}
\label{fig:elec_ncr}
\end{figure}

\begin{figure}[H]
\centering
\includegraphics[width=0.48\textwidth]{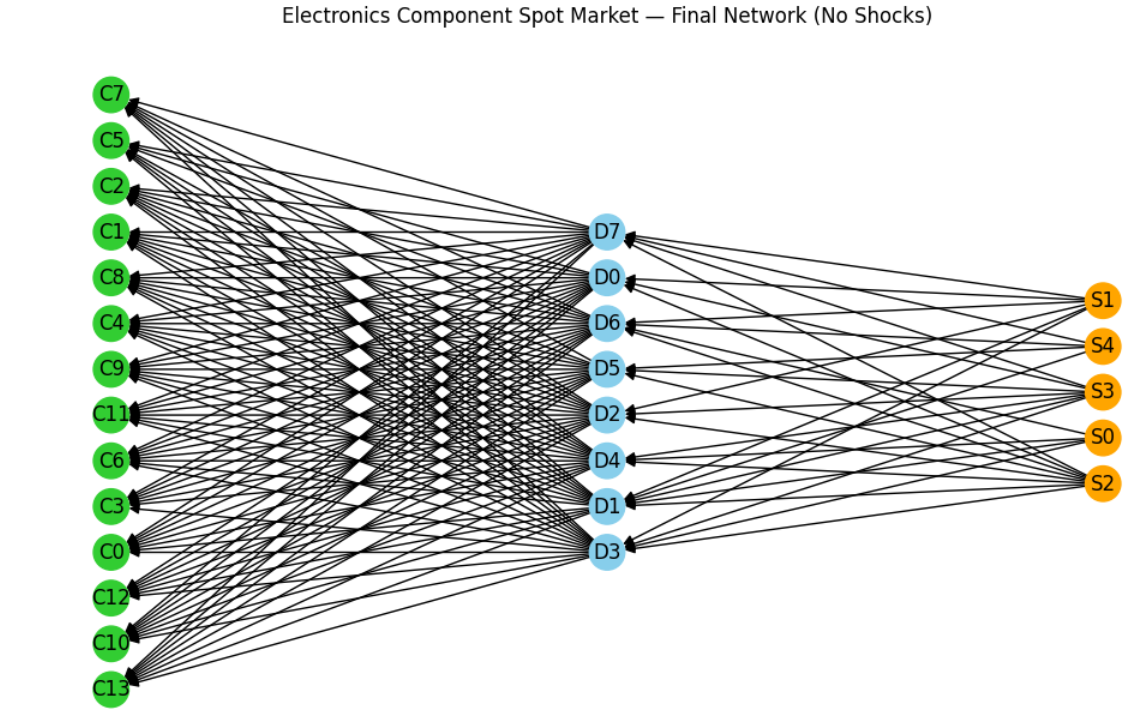}
\includegraphics[width=0.48\textwidth]{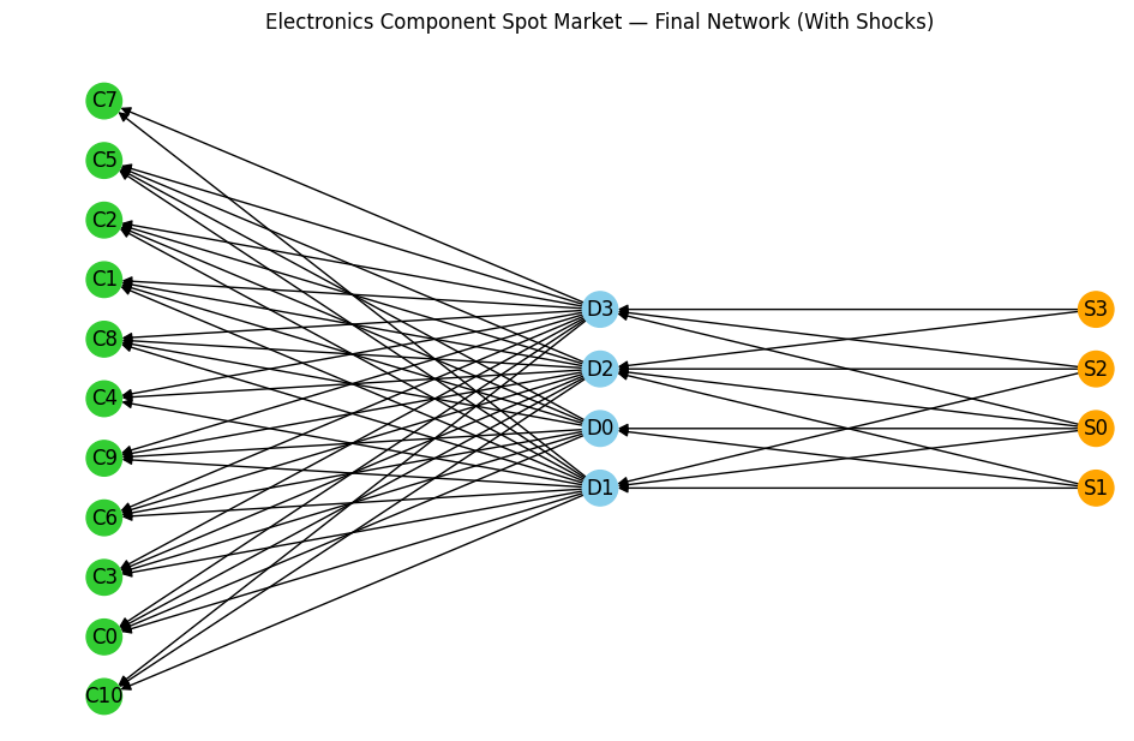}
\caption{Electronics spot market: final networks under both scenarios.}
\label{fig:elec_networks}
\end{figure}

The electronics market displays a selective consolidation mechanism. High price volatility ($\sigma=0.70$) amplifies partner switching, but MLSP remains comparatively high (Figure~\ref{fig:elec_mslp}), falling only marginally under shocks. This reflects a realistic concentration around a small number of reliable suppliers, similar to the dominance of Tier-1 semiconductor manufacturers during the 2020--2022 shortage(see \cite{semiconductor_shortage}). NCR remains low (Figure~\ref{fig:elec_ncr}), showing minimal actor exit despite significant structural reconfiguration.
The final graphs (Figure~\ref{fig:elec_networks}) illustrate the emergence of star-like formations around core suppliers. This outcome agrees with industry evidence of upstream consolidation (see \cite{sia_supply_chain}) and supports the high-volatility regime predictions of Theorem~\ref{thm:volatility}, where persistent relationships survive primarily by concentrating around stable hubs.

\subsection{Perishable Agricultural Goods}
\label{subsec:perishables}

\begin{figure}[H]
\centering
\includegraphics[width=0.48\textwidth]{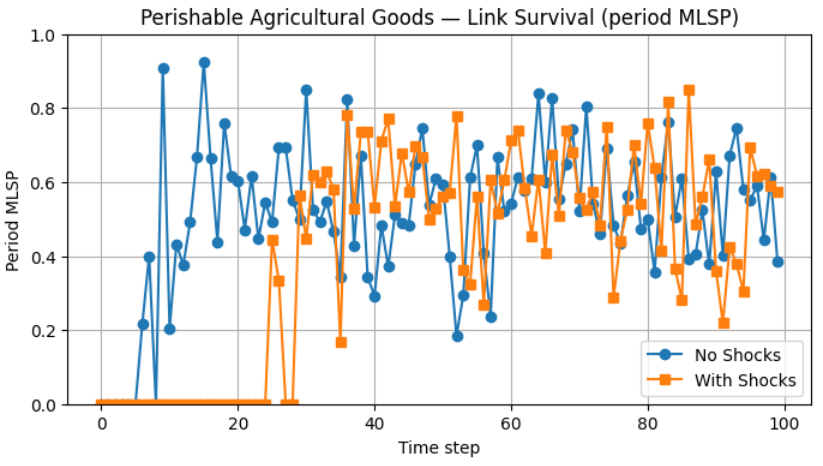}
\includegraphics[width=0.48\textwidth]{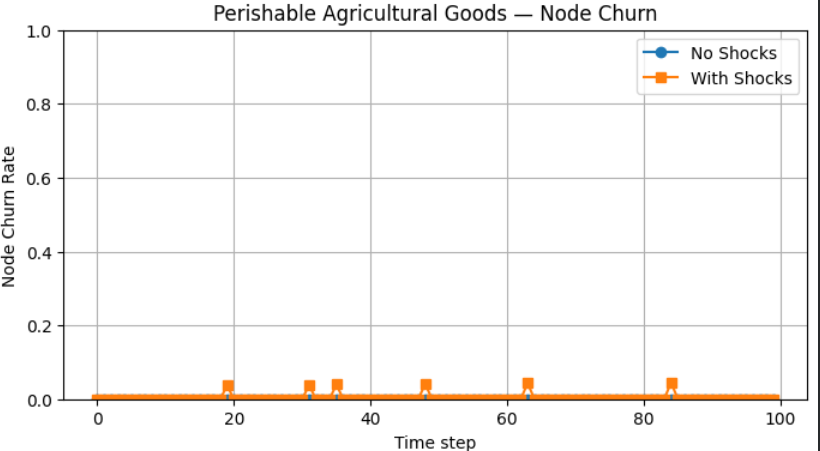}
\caption{Perishable goods: MLSP and NCR across 100 periods.}
\label{fig:perish_mslp}
\label{fig:perish_ncr}
\end{figure}

\begin{figure}[H]
\centering
\includegraphics[width=0.48\textwidth]{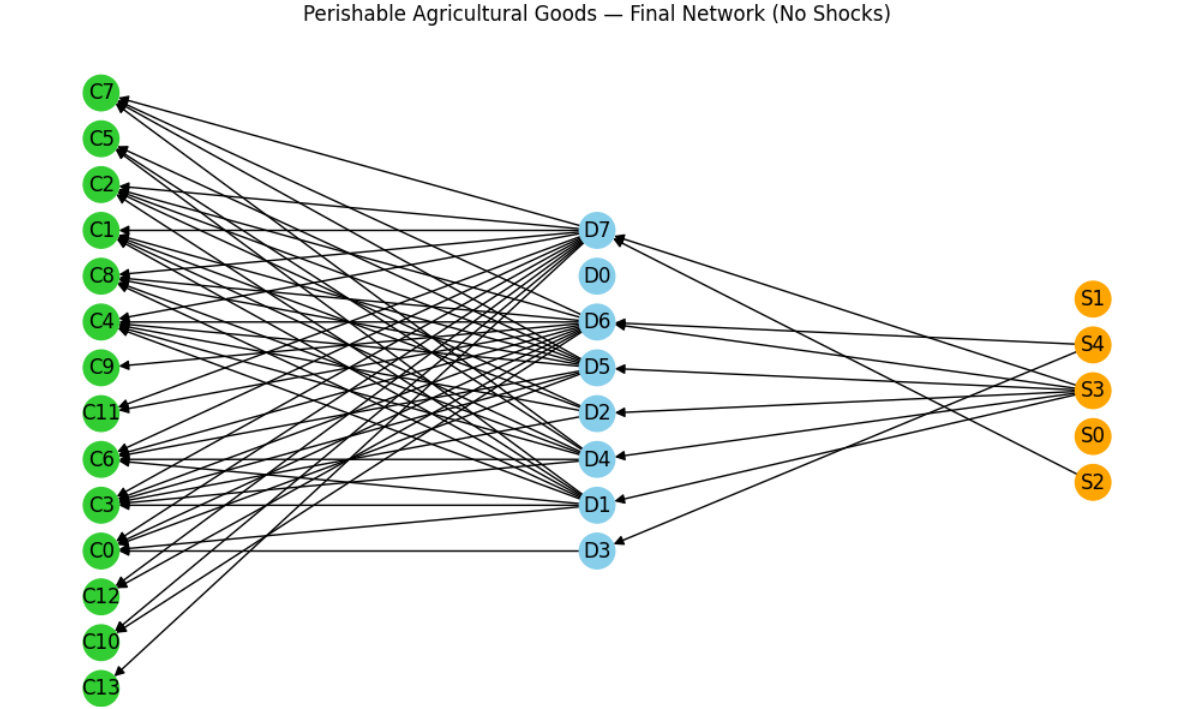}
\includegraphics[width=0.48\textwidth]{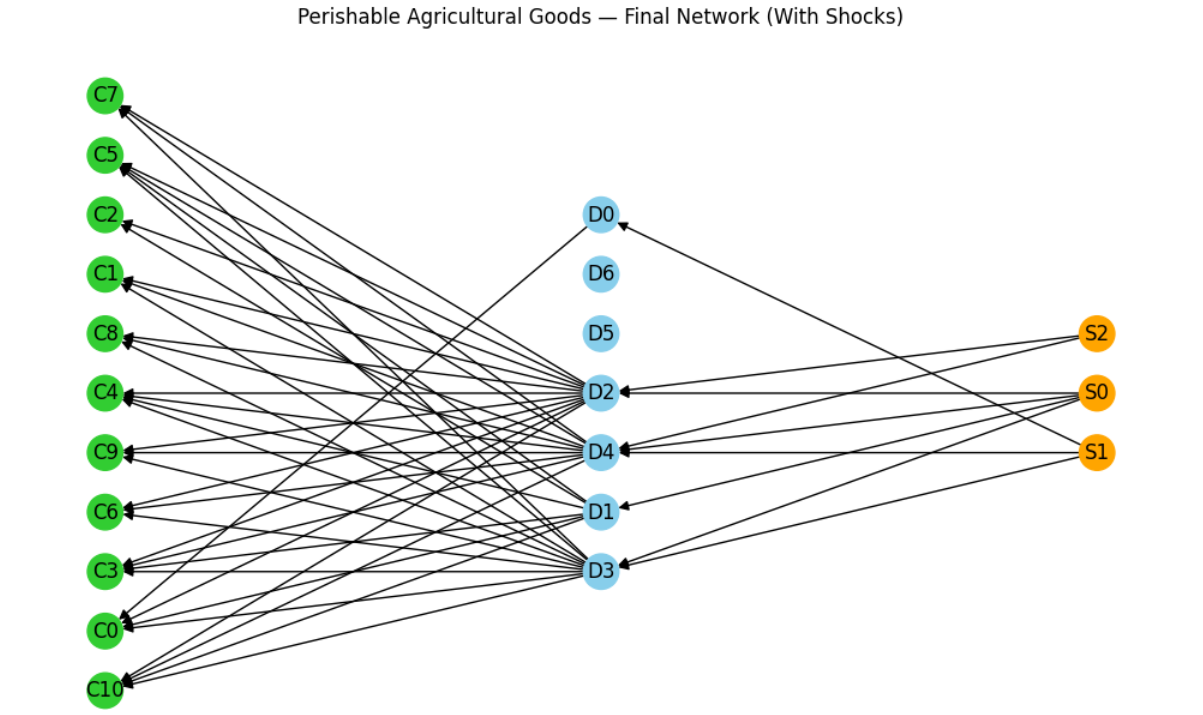}
\caption{Perishable goods: final networks under baseline and shock scenarios.}
\label{fig:perish_networks}
\end{figure}

Perishables exhibit the lowest structural robustness. In the baseline case, MLSP fluctuates moderately, but shocks produce a sharp decline (Figure~\ref{fig:perish_mslp}), consistent with the high sensitivity to trust loss in fresh-produce logistics. NCR remains low (Figure~\ref{fig:perish_ncr}), indicating that firms continue to participate even when relationships break down.
The final networks (Figure~\ref{fig:perish_networks}) are sparse, reflecting fragmentation driven by spoilage (modeled via $\gamma=0.90$) and price instability. These outcomes replicate empirical evidence from agri-food supply chains, where link failures are more frequent than firm exits (see \cite{oecd_agri_volatility,fao_food_volatility}). The model’s behavior aligns with theory; perishability and volatility jointly push the system toward the unstable regime when $\sigma$ approaches the critical threshold.

\subsection{Comparative Analysis and Validation}
\label{subsec:comparative_validation}
Across all three industries, the simulation results are consistent with the theoretical predictions. Fast fashion remains adaptively stable under moderate volatility; electronics exhibits consolidation around core suppliers under high volatility; and perishables display pronounced fragility under shocks while maintaining actor participation. The joint behavior of MLSP and NCR provides a useful diagnostic; a decline in MLSP without a corresponding rise in NCR indicates continued functional operation but increasing structural brittleness. These patterns are consistent with documented industry behaviors, suggesting that the developed framework captures key qualitative features of OSC dynamics while preserving analytical coherence.
\subsection{Managerial Insights}
\label{subsec:managerial_insights}

The simulations show that instability in OSCs stems mainly from relationship breakdown rather than firm exit.We show that across industries, firms usually stay involved even after shocks, but the Mean Link Survival Probability (MLSP) drops sharply in perishables. So, a falling MLSP is an early warning sign of structural weakness, even when the number of firms looks stable.

Fast-fashion supply chains maintain stability through diversified yet trust-based portfolios. When volatility remains below the critical threshold (Theorem~\ref{thm:volatility}), trust dampens switching disruptions, enabling flexibility without sacrificing continuity. In contrast, electronics supply chains respond to high volatility by consolidating around a few reliable suppliers, enhancing short-run stability but increasing dependency risk and the need for backup capacity.

Perishable supply chains are the most vulnerable. Volatility combined with product decay accelerates link dissolution without reducing firm participation, indicating that market adjustment alone is insufficient; coordination, shorter lead times, and preservation investments are essential.

Overall, resilience depends not only on volatility levels but on how volatility interacts with trust and product characteristics. Monitoring relational stability alongside participation metrics and aligning governance with industry-specific risks can help design sourcing strategies that remain adaptive without becoming fragile.

\section{ Conclusion } \label{sec:conclusion}

This study develops an integrated GBM–Bayesian–LOLOG framework to analyze the stability of opportunistic supply chains (OSCs) under stochastic pricing, adaptive trust, and endogenous network evolution. The analytical results establish locally stable network configurations and identify a critical volatility threshold beyond which relational persistence declines and fragmentation accelerates. Computational experiments illustrate these mechanisms in qualitatively realistic OSC environments, showing strong consistency between the theoretical predictions and industry-observed behavior.
Agent-based simulations for fast fashion, electronics, and perishable supply chains reveal that adaptive rewiring around trusted partners in fast fashion, volatility-driven consolidation in electronics, and heightened fragility in perishables. Across all settings, instability arises primarily through relationship dissolution rather than firm exit.
Overall, the findings suggest that OSCs can sustain conditional stability despite decentralized decision-making and short-term contracting. Future work may extend the framework to multi-layer and cross-border systems, explore alternative stochastic price processes, and pursue empirical calibration using transaction-level and digital trace data.

\section*{APPENDIX A }
\section*{Bayesian Belief Updating}

This appendix provides the theoretical foundation underlying the belief update rule used in Subsection \ref{sec_belief_update}.

\subsection*{Bayesian Learning Framework}

Let $B_{ij}(t) \in [0,1]$ denote agent $i$'s belief at time $t$ that partner $j$ is reliable. After each transaction, agent $i$ observes a binary outcome $s_{ij}(t) \in \{0,1\}$, where $1$ indicates successful fulfillment and $0$ indicates failure. Beliefs are updated according to Bayes' rule:
\begin{equation}\label{eq:bayes_update_app}
B_{ij}(t{+}1)
=
\frac{P(s_{ij}(t)\mid \theta_j)\,B_{ij}(t)}
{P(s_{ij}(t)\mid \theta_j)\,B_{ij}(t)
+
P(s_{ij}(t)\mid \neg\theta_j)\,(1-B_{ij}(t))},
\end{equation}
where $\theta_j$ denotes the hypothesis that agent $j$ behaves cooperatively.

\subsection*{Recursive and Log-Odds Representation}

For sequential updating, beliefs may be expressed recursively across behavioral types. Let $H_j$ denote the latent type of agent $j$, and let $B_{ij}^h(t)$ be the belief that $j$ is of type $h$ at time $t$. The recursive update is
\begin{equation}
B_{ij}(t{+}1)
=
\frac{P(s_{ij}(t{+}1)\mid H_j)\,B_{ij}(t)}
{\sum_h P(s_{ij}(t{+}1)\mid H_j{=}h)\,B_{ij}^h(t)}.
\end{equation}

For numerical stability, beliefs can be expressed in log-odds form:
\begin{equation}
\tau_{ij}(t) = \frac{B_{ij}(t)}{1-B_{ij}(t)},
\end{equation}
which yields the additive update
\begin{equation}
\log \tau_{ij}(t{+}1)
=
\log \tau_{ij}(t)
+
\log \Lambda_{ij}(t{+}1),
\end{equation}
where $\Lambda_{ij}(t{+}1)$ is the likelihood ratio associated with the observed outcome.

\subsection*{Smoothed Update as a Bayesian Approximation}

While exact Bayesian updating is theoretically appealing, it becomes computationally burdensome in large-scale, dynamically evolving networks. The smoothed belief update used in the main text,
\[
B_{ij}(t{+}1) = (1-\lambda)\,B_{ij}(t) + \lambda\,s_{ij}(t{+}1),
\]
can be interpreted as an approximation to Bayesian learning under a conjugate Beta prior with exponential forgetting. This exponentially weighted moving average preserves the qualitative behavior of Bayesian updating while remaining tractable in simulations involving frequent interactions and network reconfiguration.

\section*{APPENDIX B}
\textbf{Proof of Theorem \ref{thm:stable_network} : }
The proof proceeds in four steps using Kakutani’s fixed-point theorem (see \cite{Kakutani1941}).

\textit{Step 1 (Convexification):}
By Assumption~\textbf{A.1}, the set of feasible network configurations 
\(\mathcal{G}\subseteq\{0,1\}^{|V|\times|V|}\) is finite and nonempty.  
Hence its convex hull,
\[
\operatorname{conv}(\mathcal{G})
=\Big\{\sum_{k} \lambda_k G_k : G_k \in \mathcal{G},\ 
\lambda_k \ge 0,\ \sum_k \lambda_k=1\Big\},
\]
is compact and convex in \(\mathbb{R}^{|V|\times|V|}\).  
Elements of \(\operatorname{conv}(\mathcal{G})\) are interpreted as 
\emph{probabilistic link configurations}, where \(g_{ij}\in[0,1]\) 
represents the probability or intensity of link \((i,j)\).  
Compactness and convexity here follow directly from the finiteness of 
\(\mathcal{G}\) (Assumption~\textbf{A.1}) and the boundedness of the 
feasible strategy space.

\textit{Step 2 (Best-response correspondence):}
For each configuration \(G \in \operatorname{conv}(\mathcal{G})\), 
define the best-response correspondence
\[
\Phi(G)
=\Big\{\, G' \in \operatorname{conv}(\mathcal{G}) :\ 
g'_{ij}>0 \Rightarrow \mathbb{E}[\pi_j^t(i;G)] \ge \varepsilon \Big\}.
\]
The expected-profit function \(\mathbb{E}[\pi_j^t(i;G)]\) is continuous 
in \(G\), in the stochastic prices (GBM), and in belief parameters 
by Assumption~\textbf{S.1} together with the global smoothness and 
integrability conditions in Assumptions~\textbf{A.2--A.3}.  
Therefore, \(\Phi(G)\) is well defined and measurable.  
\(\Phi(G)\) can be interpereted as set of all configurations 
in which every active link is at least weakly profitable 
relative to the incumbent network~\(G\).

\textit{Step 3 (Verification of Kakutani’s conditions):}
We now verify that \(\Phi\) satisfies the four requirements 
of Kakutani’s fixed-point theorem.

\begin{itemize}
    \item \emph{Nonemptiness:}  
    By Assumption~\textbf{S.2} (profitability threshold) 
    and the global feasibility of profitable links 
    from Assumption~\textbf{A.4}, 
    there exists at least one configuration \(G^\# \in \mathcal{G}\)
    such that all active links satisfy 
    \(\mathbb{E}[\pi_j^t(i;G^\#)] > \varepsilon\).  
    Hence, for any incumbent \(G\),
    the set \(\Phi(G)\) contains \(G^\#\), and therefore 
    \(\Phi(G)\neq\emptyset\).
    
    \item \emph{Convexity:}  
Let \(G_1', G_2' \in \Phi(G)\) and let \(\lambda \in [0,1]\). 
Consider the convex combination 
\(G'_\lambda = \lambda G_1' + (1-\lambda) G_2'\).  
By definition of \(\Phi(G)\), every link \((i,j)\) with 
\(g'_{1,ij}>0\) or \(g'_{2,ij}>0\) satisfies 
\(\mathbb{E}[\pi_j^t(i;G)] \ge \varepsilon\) 
for the fixed incumbent configuration \(G\).  
If \((G'_\lambda)_{ij} > 0\), then at least one of 
\(g'_{1,ij}\) or \(g'_{2,ij}\) must be positive, 
and hence the same profitability condition holds for that link.  As expected profits are evaluated at the fixed \(G\) 
(and are continuous in \(G\) by Assumption~\textbf{S.1}), 
this implication is preserved for all convex combinations.  
Therefore, \(G'_\lambda \in \Phi(G)\), 
and thus the correspondence \(\Phi(G)\) has convex values.
    
    \item \emph{Compactness:}  
Since \(\Phi(G)\subseteq\operatorname{conv}(\mathcal{G})\) 
and \(\operatorname{conv}(\mathcal{G})\) is compact by Step~1, 
each value \(\Phi(G)\) is compact.  
This property follows directly from the finiteness of the feasible set 
\(\mathcal{G}\) ensured by Assumption~\textbf{A.1}. Now as each adjacency matrix \(G_k\) has entries in \([0,1]\), 
the feasible set is bounded and closed in 
\(\mathbb{R}^{|V|\times|V|}\); therefore its convex hull is also 
closed and bounded, and thus compact by the Heine--Borel theorem (\cite{rudin1976}).
    
    \item \emph{Upper hemicontinuity:}  
   By Assumption~\textbf{S.1}, the mapping 
\(G \mapsto \mathbb{E}[\pi_j^t(i;G)]\) is continuous for all 
links \((i,j)\).  
For each link, consider the set
\[
A_{ij}
:= 
\big\{(G,G') \in 
\operatorname{conv}(\mathcal{G}) \times \operatorname{conv}(\mathcal{G})
:\ g'_{ij} \le 0 
\ \text{or}\ 
\mathbb{E}[\pi_j^t(i;G)] \ge \varepsilon \big\}.
\]
Since \(G \mapsto \mathbb{E}[\pi_j^t(i;G)]\) is continuous, the set
\(\{(G,G'):\mathbb{E}[\pi_j^t(i;G)] \ge \varepsilon\}\) is closed, and so is
\(\{(G,G') : g'_{ij} \le 0\}\); their union \(A_{ij}\) is therefore closed.  
The graph of the correspondence \(\Phi\) can be written as
\[
\operatorname{Graph}(\Phi)
= \bigcap_{(i,j)} A_{ij},
\]
which is a finite intersection of closed sets and hence closed.  
Because each value \(\Phi(G)\) is compact as established above, 
the closed-graph property implies that 
\(\Phi\) is upper hemicontinuous on 
\(\operatorname{conv}(\mathcal{G})\).
\end{itemize}

\textit{Step 4 (Fixed-point existence ):}
All the conditions of Kakutani’s fixed-point theorem
\cite{Kakutani1941} are now satisfied.  
The domain $\operatorname{conv}(\mathcal{G})$ is nonempty, compact, and convex
(by Assumption~\textbf{A.1});  
and the correspondence $\Phi(G)$ is nonempty (by Assumption~\textbf{S.2}),  
convex-valued, compact-valued, and upper hemicontinuous
(Steps~2--3 together with Assumptions~\textbf{A.1--A.2} and~\textbf{S.1}).  
Hence, by Kakutani’s theorem, there exists a fixed point  
$G^{*}\in\operatorname{conv}(\mathcal{G})$ such that $G^{*}\in\Phi(G^{*})$.

For every link $(i,j)$ with $g^{*}_{ij}>0$, the definition of $\Phi$
implies $\mathbb{E}[\pi_j^t(i;G^{*})]\ge\varepsilon$.
Because $\operatorname{conv}(\mathcal{G})$ is the convex hull of the finitely many
pure configurations in $\mathcal{G}$, its extreme points are exactly the
elements of $\mathcal{G}$.
By the   Krein--Milman theorem, any nonempty compact convex set such as
$\Phi(G^{*})$ contains at least one extreme point.
Let $\tilde G$ denote such an extreme point of $\Phi(G^{*})$;
then necessarily $\tilde G\in\mathcal{G}$, i.e., $\tilde G$ is a
\emph{pure network configuration}.  
Moreover, by construction, $\tilde G\in\Phi(G^{*})$,
so every active link in $\tilde G$ satisfies
$\mathbb{E}[\pi_j^t(i;G^{*})]\ge\varepsilon$.
Continuity of $\mathbb{E}[\pi_j^t(i;G)]$ in $G$
(Assumption~\textbf{S.1}) and finiteness of $\mathcal{G}$
ensure that expected profits vary continuously across configurations; thus
the same inequality holds for $\tilde G$ when profits are evaluated at
$\tilde G$ itself.

By the dissolution rule in Assumption~\textbf{S.2},
no agent can profitably deviate by forming links whose expected profit
falls below~$\varepsilon$.
Therefore, $\tilde G$ constitutes a \emph{locally stable network configuration}.  
Under Assumptions~\textbf{A.1--A.4} and~\textbf{S.1--S.2},
this pure configuration $\tilde G$ (and the associated mixed representation
$G^{*}$) represents a self-consistent equilibrium in which all active links
are profitable and no unilateral deviation increases expected profit. \qed

\textbf{ Proof of Theorem \ref{thm:volatility}:}
By Assumptions~\textbf{B.2}–\textbf{B.3}, for each fixed \(G\) the map
\(\sigma\mapsto U_{ij}(\sigma;G)\) is continuous and nonincreasing. Since
\(F\) is continuous and strictly increasing (Assumption~\textbf{B.1}), the
composition \(\pi_{ij}(\sigma;G)=F(U_{ij}(\sigma;G))\) is continuous and
nonincreasing in \(\sigma\). By Assumption~\textbf{B.4} the sets
\(\mathcal{N}(G)\) are finite, hence \(S(\sigma)\) is a finite average of
continuous, nonincreasing functions. Therefore \(S(\sigma)\) is continuous
and nonincreasing on \([0,\infty)\).

Because \(0\le S(\sigma)\le 1\) for all \(\sigma\) and \(S(\cdot)\) is
nonincreasing, the limit \(S_\infty := \lim_{\sigma\to\infty} S(\sigma)\)
exists and is finite. If \(S(0)>S_\infty\), choose any \(s^*\) with
\(S_\infty < s^* < S(0)\). Continuity of \(S\) guarantees the existence of
at least one \(\sigma\ge0\) with \(S(\sigma)=s^*\). Define
\[
\sigma_c := \inf\{\sigma\ge0:\ S(\sigma)<s^*\}.
\]
By definition of the infimum and the monotonicity of \(S\), we have
\(S(\sigma)\ge s^*\) for all \(\sigma<\sigma_c\), while for all
\(\sigma>\sigma_c\) values of \(S\) lie below \(s^*\) (values immediately
above the infimum fall below \(s^*\) by construction). If the infimum is
attained then \(S(\sigma_c)=s^*\); otherwise \(S(\sigma_c)\ge s^*\) and
strict inequality holds for all sufficiently large \(\sigma>\sigma_c\).
This establishes the stated threshold behaviour.\qed

\textbf{Proof of Theorem \ref{thm:node_importance}:}
By definition,
\[
I_i(G)=\sum_{j:(i,j)\in E}\mathbb{E}[\pi_j^t(i)]\,B_{ij}(t),
\]
a finite sum of nonnegative terms. Under Assumptions~\textbf{A.2--A.3} and the trust-update rule~\eqref{eq:smoothing_update} each factor is continuous; hence \(I_i(G)\) is continuous and coordinate-wise nondecreasing in the collection \(\{\mathbb{E}[\pi_j^t(i)],B_{ij}(t)\}_{j:(i,j)\in E}\).

By Assumption~C.1 the link probability \(\pi_{ij}(\sigma)\) is continuously differentiable in \(\mathbb{E}[\pi_j^t(i)]\) and \(B_{ij}(t)\), with strictly positive partial derivatives. From Theorem~\ref{thm:volatility} the network stability function is
\[
S(\sigma)=\frac{1}{|E|}\sum_{(k,\ell)\in E}\pi_{k\ell}(\sigma),
\]
so each term \(\pi_{ij}\) depends smoothly on expected profitability and trust.

Consider a small component-wise nonnegative perturbation
$$\Delta_i=\{(\Delta\mathbb{E}[\pi_j^t(i)],\Delta B_{ij})\}_{j:(i,j)\in E},$$ not all zero. By the differentiability of the \(\pi_{ij}\)'s and finiteness of the sum, the first-order (directional) change in \(S\) equals the linear term plus a higher-order remainder:
\[
\Delta S(\sigma)
=\sum_{j:(i,j)\in E}\left(
\frac{\partial \pi_{ij}}{\partial \mathbb{E}[\pi_j^t(i)]}\,\Delta\mathbb{E}[\pi_j^t(i)]
+\frac{\partial \pi_{ij}}{\partial B_{ij}(t)}\,\Delta B_{ij}
\right) + o(\|\Delta_i\|).
\]
Each partial derivative is nonnegative (strictly positive under C.1) and at least one component of \(\Delta_i\) is strictly positive; therefore the leading linear term is strictly positive. For sufficiently small perturbations the remainder \(o(\|\Delta_i\|)\) is negligible, hence \(\Delta S(\sigma)>0\). Equivalently, the directional derivative of \(S\) in the direction of \(\Delta_i\) is positive, so \(\partial S(\sigma)/\partial I_i(G)>0\) along that direction. This shows that increases in \(I_i(G)\) raise the network stability measure \(S(\sigma)\), completing the proof.
\qed

\textbf{Proof of Theorem\ref{thm:procurement_utility}:}
By Assumption~\textbf{A.2}, each resale price \(P_{jk}(\cdot)\) is 
\(C^2\), strictly decreasing (\(P'_{jk}<0\)), and concave (\(P''_{jk}\le0\)).
Let \(r_{jk}(q)=P_{jk}(q)q\).
Then 
\(r_{jk}''(q)=P_{jk}''(q)q+2P_{jk}'(q)<0\),
so \(r_{jk}\) is strictly concave.  
Since \(B_{jk}\ge0\), the mapping 
\(q\mapsto\mathbb{E}[B_{jk}P_{jk}(q)q]\) 
is also strictly concave.

From Assumption~\textbf{A.3}, 
\(c_i(\cdot)\) is \(C^1\) with \(c_i'(q)\ge0\) and \(c_i''(q)\ge0\).  
Defining \(g(q)=c_i(q)q\) with $c_i(q)=p_i^t-\beta_i(q)$ gives 
\(g''(q)=c_i''(q)q+2c_i'(q)\ge0\),
so \(g (\cdot)\) is convex.

Hence, the objective function
\[
\mathcal{U}_j(q)
=\sum_{k\in\mathcal{K}_j}\mathbb{E}[B_{jk}P_{jk}(q)q]
-c_i(q)q
\]
is the sum of strictly concave terms minus a convex term,
therefore strictly concave on its domain.  
By Assumption~\textbf{A.4}, the feasible set \([0,q^{\max}]\) is compact;
a continuous, strictly concave function on a compact interval
has a unique maximizer, denoted \(q^\ast\).

If \(q^\ast\) lies in the interior of the interval, 
the necessary and sufficient condition for optimality is 
\(\mathcal{U}_j'(q^\ast)=0\).
Differentiation under the expectation is justified 
by the regularity and integrability in Assumption~\textbf{A.2}, giving
\[
\frac{d}{dq}\,\mathbb{E}[B_{jk}P_{jk}(q)q]
=\mathbb{E}\!\left[B_{jk}\,\frac{d}{dq}(P_{jk}(q)q)\right].
\]
Substituting this expression and differentiating \(c_i(q)q\)
yields the stated first-order condition.
If \(q^\ast\) lies on the boundary 
(\(q^\ast\in\{0,q^{\max}\}\)), 
the global maximum occurs at that point by strict concavity.
\qed

\textbf{ Proof of Theorem \ref{thm:comparative}:}
Define
\begin{equation}
F(q;B,p_i^t)
:= \sum_{k}
\frac{\partial}{\partial q}\,\mathbb{E}[B_{jk}P_{jk}(q)q]
-\big(c_i'(q)q + c_i(q)\big),
\label{eq:def_F}
\end{equation}
so that the first-order condition for an interior maximizer can be written as
\begin{equation}
F(q^\ast;B,p_i^t)=0.
\label{eq:FOC_interior}
\end{equation}

By Theorem~\ref{thm:procurement_utility} and Assumptions~\textbf{A.2--A.4},  
the function $F$ is continuously differentiable in all its arguments.  
Differentiating~\eqref{eq:def_F} with respect to $q$ yields
\begin{equation}
F_q(q;B,p_i^t)
=\sum_k \frac{\partial^2}{\partial q^2}\mathbb{E}[B_{jk}P_{jk}(q)q]
-\big(c_i''(q)q + 2c_i'(q)\big).
\label{eq:Fq}
\end{equation}
Because each expected revenue term $\mathbb{E}[B_{jk}P_{jk}(q)q]$ is strictly concave in $q$
and $c_i''(q),c_i'(q)\ge0$, we have $F_q(q;B,p_i^t)<0$.  
This strict negativity ensures that $F$ is strictly decreasing in $q$, so the implicit function theorem applies, implying that the optimal procurement quantity $q^\ast(B,p_i^t)$ is differentiable in both parameters $B$ and $p_i^t$.

Next, we compute the partial derivatives of $F$ with respect to $B_{jk}$ and $p_i^t$.  
Because $B_{jk}$ enters linearly in the expected term and $c_i(q)=p_i^t-\beta_i(q)$, we obtain
\begin{align}
F_{B_{jk}}(q;B,p_i^t)
&= \frac{\partial}{\partial B_{jk}}F(q;B,p_i^t)
= \frac{\partial}{\partial q}\mathbb{E}[P_{jk}(q)q],
\label{eq:FBjk}\\[3pt]
F_{p_i^t}(q;B,p_i^t)
&= -\frac{\partial}{\partial p_i^t}\big(c_i'(q)q+c_i(q)\big)
= -1.
\label{eq:Fpi}
\end{align}

Applying the implicit function theorem to~\eqref{eq:FOC_interior}, the derivatives of $q^\ast$ with respect to $B_{jk}$ and $p_i^t$ are
\begin{equation}
\frac{\partial q^\ast}{\partial B_{jk}}
= -\frac{F_{B_{jk}}(q^\ast;B,p_i^t)}{F_q(q^\ast;B,p_i^t)},
\qquad
\frac{\partial q^\ast}{\partial p_i^t}
= -\frac{F_{p_i^t}(q^\ast;B,p_i^t)}{F_q(q^\ast;B,p_i^t)}.
\label{eq:partials}
\end{equation}

To study the effect of supplier price, note from~\eqref{eq:Fpi} that $F_{p_i^t}=-1$.  
Substituting into~\eqref{eq:partials} gives
\[
\frac{\partial q^\ast}{\partial p_i^t}
= -\frac{-1}{F_q(q^\ast;B,p_i^t)}
= \frac{1}{F_q(q^\ast;B,p_i^t)}.
\]
Since $F_q(q^\ast;B,p_i^t)<0$, it follows that
\begin{equation}
\frac{\partial q^\ast}{\partial p_i^t} < 0,
\label{eq:price_effect}
\end{equation}
confirming that the optimal procurement quantity decreases when the supplier price increases.

Next, consider the comparative static with respect to the trust belief $B_{jk}$.  
From~\eqref{eq:FBjk} and~\eqref{eq:partials},
\begin{equation}
\frac{\partial q^\ast}{\partial B_{jk}}
= -\frac{\dfrac{\partial}{\partial q}\mathbb{E}[P_{jk}(q)q]\big|_{q=q^\ast}}
{F_q(q^\ast;B,p_i^t)}.
\label{eq:belief_effect}
\end{equation}
Because $F_q(q^\ast;B,p_i^t)<0$, the sign of $\partial q^\ast/\partial B_{jk}$ 
matches the sign of the marginal revenue term 
$\dfrac{\partial}{\partial q}\mathbb{E}[P_{jk}(q)q]\big|_{q=q^\ast}$.  
Thus, whenever the expected marginal revenue is nonnegative at the optimum,
\begin{equation}
\frac{\partial q^\ast}{\partial B_{jk}} \ge 0,
\label{eq:belief_sign}
\end{equation}
and the optimal procurement quantity is nondecreasing in the belief $B_{jk}$.
This condition holds under standard demand specifications where expected price or sales probability is nonincreasing in $q$.\footnote{%
The monotonicity with respect to $B_{jk}$ holds under the mild local condition 
$\frac{d}{dq}\mathbb{E}[P_{jk}(q)q]\big|_{q=q^\ast}\ge0$, 
which ensures that increased trust enhances expected marginal revenue at the optimum.}
Equations~\eqref{eq:price_effect} and~\eqref{eq:belief_sign} together establish that 
$q^\ast$ decreases with supplier price and increases with downstream trust, 
thus proving parts (ii) and (i) of the theorem, respectively.
\qed

\textbf{ Proof of Theorem \ref{thm:ergodicity}:}
Under Assumption~\textbf{A.1}, the feasible state space 
$\mathcal{G}$ of OSC network configurations is finite.  
The LOLOG specification in~\eqref{eqn_LOLOG} defines a 
time-homogeneous Markov chain $\{G_t\}_{t\ge0}$ on $\mathcal{G}$, 
with transition probabilities $P(G \to G')$.  
A \emph{feasible single-edge toggle} refers to the addition or deletion 
of a single link that respects the feasibility constraints of the OSC model 
(e.g., capacity, reciprocity, or institutional restrictions).

\textit{Irreducibility:}
Let $G,G'\in\mathcal{G}$.  
The symmetric difference of their edge sets, 
$\Delta(G,G') = E(G)\triangle E(G')$,  
is finite.  
By successively toggling each edge in $\Delta(G,G')$ one at a time, 
we obtain a finite sequence of feasible transitions
\[
G = G^{(0)} \to G^{(1)} \to \cdots \to G^{(m)} = G',
\]
each occurring with probability at least $\eta>0$ by the theorem’s assumption.  
Therefore, there exists a path from $G$ to $G'$ with strictly positive probability, 
and the chain is \emph{irreducible}.

\textit{Aperiodicity:}
For every $G\in\mathcal{G}$, the self-transition probability satisfies 
$P(G\to G)\ge\eta>0$.  
Because each state can transition to itself with positive probability, 
its period equals one, and hence the chain is \emph{aperiodic}.

\textit{Existence and uniqueness of the stationary distribution:}
Since $\{G_t\}$ is a finite, irreducible, and aperiodic Markov chain, 
standard results in Markov chain theory 
(see \cite[Thm.~1.7.7]{norris1997markov}) imply the existence of 
a unique stationary distribution $\pi^\ast$ on $\mathcal{G}$ satisfying 
$\pi^\ast = \pi^\ast P$.  
Moreover, for any initial distribution $\pi_0$, 
we have $\pi_0 P^t \to \pi^\ast$ as $t\to\infty$.

\textit{Ergodic convergence:}
By the strong law of large numbers for finite-state ergodic Markov chains 
(see \cite[Thm.~1.10.2]{norris1997markov}),  
for any bounded measurable function $f:\mathcal{G}\to\mathbb{R}$,
\[
\frac{1}{T}\sum_{t=1}^{T} f(G_t)
\xrightarrow{\text{a.s.}}
\sum_{G\in\mathcal{G}} f(G)\,\pi^\ast(G)
\qquad \text{as } T \to \infty,
\]
which establishes the ergodic convergence result in~\eqref{eq:ergodic_convergence}.

\qed

\bibliographystyle{elsarticle-num}
\bibliography{ref}

\end{document}